\documentclass[lettersize,journal]{IEEEtran}
\usepackage{amsmath,amsfonts}
\usepackage{algorithmic}
\usepackage{algorithm}
\usepackage{array}
\usepackage[caption=false,font=normalsize,labelfont=sf,textfont=sf]{subfig}
\usepackage{textcomp}
\usepackage{stfloats}
\usepackage{url}
\usepackage{verbatim}
\usepackage{graphicx}
\usepackage{cite}
\usepackage{amssymb}
\usepackage{hyperref}
\usepackage{tikz}
\usepackage{pgfplots}
\pgfplotsset{compat=1.16}
\usepgfplotslibrary{groupplots}
\pgfplotsset{
  every axis/.append style={
    no markers,
    grid=major,
    grid style={dashed},
    legend style={font=\tiny},
    ylabel style={font=\scriptsize},
    xlabel style={font=\scriptsize},
  },
  every axis plot/.append style={line width=1.2pt, line join=round},
  every axis legend/.append style={legend columns=1},
  group/group size=3 by 1,
  every x tick label/.append style={alias=XTick,inner xsep=0pt},
  every x tick scale label/.style={at=(XTick.base east),anchor=base west}
}

\definecolor{col1}{RGB}{53, 110, 175}
\definecolor{col2}{RGB}{204, 42, 42}
\definecolor{col3}{RGB}{255, 175, 35}
\definecolor{col4}{RGB}{79, 162, 46}
\definecolor{col5}{RGB}{97, 97, 97}
\definecolor{col6}{RGB}{103, 63, 153}
\definecolor{col7}{RGB}{0, 0, 0}
\definecolor{col8}{RGB}{123, 63, 0}

\tikzset{
  curve1/.style={col1},
  curve2/.style={col2},
  curve3/.style={col4},
  curve4/.style={col3},
  curve5/.style={col6},
  curve6/.style={cyan},
  curve7/.style={col5, dashdotted},
  curve8/.style={col7, dashed},
  curve9/.style={col8, densely dotted},
  curve10/.style={teal, densely dotted},
  curve11/.style={lime},
  curve12/.style={orange},
}
\newcommand{\discreteVectorField}{\mathcal{V}}
\newcommand{\numberProcesses}{n_p}

\newcommand{\domain}{\mathcal{M}}

\newcommand{\simplex}{\sigma}

\newcommand{\diagram}{\mathcal{D}}

\newcommand{\lt}{<}

\newcommand{\boundary}[1]{\emph{Boundary}({#1})}
\newcommand{\localBoundary}[1]{\emph{LocalBoundary}({#1})}
\newcommand{\globalBoundary}[1]{\emph{GlobalBoundary}({#1})}
\newcommand{\globalLocalBoundary}[1]{\emph{GlobalLocalBoundary}({#1})}
\newcommand{\graph}[1]{\mathcal{G}_{{{#1}}}}
\newcommand{\localGraph}[2]{{\mathcal{G}_{{{#1}, {#2}}}}}
\newcommand{\ghostedGraph}[2]{{\mathcal{G}'_{{{#1}, {#2}}}}}
\definecolor{darkblue}{RGB}{18,22,46}
\definecolor{mediumblue}{RGB}{51,70,99}
\definecolor{lightblue}{RGB}{94,125,141}
\definecolor{verylightblue}{RGB}{185,196,189}

\definecolor{darkred}{RGB}{98,24,13}
\definecolor{orange}{RGB}{184,101,26}
\definecolor{lightyellow}{RGB}{194,171,85}
\definecolor{verylightyellow}{RGB}{200,197,132}

\newcommand{\pierre}[1]{\textcolor{black}{#1}}
\newcommand{\julien}[1]{\textcolor{black}{#1}}
\newcommand{\eve}[1]{\textcolor{black}{#1}}
\newcommand{\majorRevision}[1]{\textcolor{black}{#1}}
\newcommand{\julienRevision}[1]{\textcolor{black}{#1}}
\newcommand{\pierreRevision}[1]{\textcolor{black}{#1}}

\begin{document}

\title{Distributed Discrete Morse Sandwich: \\ 
Hybrid Distributed and Shared Memory Parallelization of Persistent Homology}

\title{Distributed Discrete Morse Sandwich: \\
Efficient Computation of Persistence Diagrams \\ 
for Massive Scalar Data
}
% Computation of Persistence
% Diagrams
% of
% Scalar Data}

\author{E. Le Guillou, P. Fortin, J. Tierny
\IEEEcompsocitemizethanks{
  \IEEEcompsocthanksitem E. Le Guillou is with the CNRS, Sorbonne Université and
  University of Lille.
  E-mail:
  \href{mailto:eve.le_guillou@sorbonne-universite.fr}{eve.le\_guillou@sorbonne-universite.fr}
  \IEEEcompsocthanksitem J. Tierny is with the CNRS and Sorbonne
  Université.
  E-mail:
  \href{mailto:julien.tierny@sorbonne-universite.fr}{julien.tierny@sorbonne-universite.fr}
  \IEEEcompsocthanksitem P. Fortin is with Univ. Lille, CNRS, Centrale
  Lille, UMR 9189 CRIStAL, F-59000 Lille, France.
  %% is with the University of Lille.
  E-mail: \href{mailto:pierre.fortin@univ-lille.fr}{pierre.fortin@univ-lille.fr}
}
}

\markboth{Journal of \LaTeX\ Class Files,~Vol.~14, No.~8, August~2021}%
{Shell \MakeLowercase{\textit{et al.}}: A Sample Article Using IEEEtran.cls for IEEE Journals}


\maketitle

\begin{abstract}
  The persistence diagram,
  \julien{which describes the topological features of a dataset, is a key
  descriptor}
  in Topological Data Analysis.
  The \emph{``Discrete Morse Sandwich''} (DMS)
  \julien{method}
%   algorithm
%   is
  \julien{has been reported to be}
%   currently
  the most
efficient \julien{algorithm for computing persistence diagrams of 3D scalar
fields on a single node,}
  using shared-memory
parallelism.
  \julien{In this work, we extend}
%   We aim here at extending
  DMS to distributed-memory parallelism
  for the efficient and scalable computation of persistence diagrams
  for massive \julien{datasets} \julien{across} multiple compute nodes.
  On \julien{the} one hand, we can leverage the embarrassingly parallel
%   feature
  \julien{procedure}
  of the first and most time-consuming step \julien{of DMS} (namely the discrete
  gradient computation).
  On the other hand, the efficient distributed computations of the
  subsequent DMS steps are much more challenging.
  \julien{To address this, we have extensively revised}
%   We have thus deeply
%   revisited
  the DMS \julien{routines}
  \julien{by contributing}
%   thanks to
  a new self-correcting \pierre{distributed}  \julien{pairing}
  algorithm, \julien{redesigning key} data structures and
  \julien{introducing}
%   to the
%   introduction
%   of
  computation tokens
%   for supervising
  \julien{to coordinate}
  distributed computations.
  We have also introduced a
%   specific
  \julien{dedicated}
  communication thread
  \julien{to overlap}
%   for
  communication and computation.
  Detailed performance analyses show the scalability of our hybrid
  MPI+thread
  \julien{approach}
%   implementation
  for strong and weak scaling using up to 16 nodes of
  32 cores
%   each
  \julien{(512 cores total).}
%   (for a total of 512 cores).
  Our
  \julien{algorithm}
%   implementation
%   also
  outperforms \majorRevision{\emph{DIPHA}},
%   \julien{to our knowledge}
%   the only public
  \julien{a reference method}
%   implementation
  for \julien{the distributed computation of persistence diagrams,}
  with an average speedup of \julien{$\times 8$} on 512 cores.
%   We provide \julien{a usage example}
%   of our \julien{open source implementation}
%   at \url{https://github.com/eve-le-guillou/DDMS-example}.
%   Finally, we show the capabilities of our
%   \julien{approach}
%   by computing the
%   persistence diagram of a \julien{public 3D scalar field} of 6 billion
% vertices in 174 seconds \julien{on} 512 cores.
  We show the practical capabilities of our
  \julien{approach}
  by computing the
  persistence diagram of a \julien{public 3D scalar field} of 6 billion
vertices in 174 seconds \julien{on} 512 cores.
  Finally, we provide \julien{a usage example}
  of our \julien{open-source implementation}
  at \url{https://github.com/eve-le-guillou/DDMS-example}.
\end{abstract}


\section{Introduction}

Modern datasets continue to grow, through constant improvements of data 
acquisition and numerical computations. Hence grows the need for efficient 
tools for the representation and analysis of those datasets. In the case 
of scalar data, \emph{Topological Data Analysis} (TDA)
% \julien{\cite{edelsbrunner09}}
has proven to
provide practical solutions for capturing and extracting structural characteristics in 
high-resolution data.
Topological methods have been used
% by the visualization community
\julien{for data analysis in several}
% to perform analysis
% and representation for
% numerous and vastly different
domains, such as
combustion \cite{laney_vis06, bremer_tvcg11, gyulassy_ev14},
fluid dynamics \cite{kasten_tvcg11, nauleau_ldav22}, nuclear energy \cite{beiNuclear16}, 
data science \cite{ChazalGOS13, topoMap}, chemistry \cite{harshChemistry, chemistry_vis14, Malgorzata19} or astrophysics \cite{sousbie11, shivashankar2016felix}.

TDA includes a wide range of representations \cite{edelsbrunner09, heine16},
several of which have been extensively studied and efficiently implemented, such
as the persistence diagram \cite{edelsbrunner02},
the Reeb graph \cite{Parsa12, gueunet_egpgv19}, the contour tree
\cite{tarasov98, carr00, gueunet_tpds19}
or the Morse-Smale complex \cite{BremerEHP04, GyulassyNPBH06, Defl15,
gyulassy_vis18}.
In this paper, we focus on
the persistence diagram \cite{edelsbrunner09}, a concise descriptor of
the topological features of a dataset\majorRevision{. This abstraction}
% \emph{topological persistence} \cite{edelsbrunner09},
% an importance metric that
helps separate noise from the noteworthy
% topological
structures of the data, \julien{enabling multi-scale data simplification
\cite{tierny_vis12, Lukasczyk_vis20},
% representation (for example with
% contour trees, Morse-Smale complexes),
segmentation \cite{carr04, topoAngler}, compression \cite{soler_pv18}, tracking
\cite{soler_ldav18, soler_ldav19, LukasczykGWBML20} or ensemble summarization
\cite{favelier_vis18, vidal_vis19, KontakVT19}.}
When considering $d$-dimensional scalar data, \julien{a persistence diagram} for
each dimension
% have
\julien{has}
to be computed, namely $\diagram_{0}$, $\diagram_{1}$ and $\diagram_{2}$
for $d=3$. Several algorithms already exist to compute these data structures, using 
either shared-memory or distributed-memory parallelism. 
In a shared-memory setting, the Discrete Morse Sandwich (DMS)
algorithm
% is proven to be currently
\julien{has been reported to be}
the most efficient implementation,
as shown in its associated benchmark \cite{guillou_tech22}.
However this setting limits the maximum size of the datasets that can be processed to the memory of 
a single node\majorRevision{. This capacity is often} insufficient with regard to the 
potentially massive size of modern datasets.

In this work, we introduce the Distributed Discrete Morse Sandwich (DDMS),
an efficient algorithm for persistence diagram computation 
exploiting distributed-memory \pierreRevision{parallel architectures}. %parallelism. 
DDMS
% is built
\julien{builds}
upon the \pierre{DMS} %Discrete Morse Sandwich (DMS)
algorithm. % \cite{guillou_tech22}.
\majorRevision{The distributed-memory \pierreRevision{parallelism}
  %configuration
  enables DDMS to leverage 
greater computational power and memory capacity, allowing a user 
to perform computation faster than DMS and on larger datasets than ever before.
The workflow of the proposed DDMS algorithm closely
\pierreRevision{follows} %parallels %% Pierre : "parallels" me semble
%introduire de la confusion vu le but du papier 
that of DMS.}
This enables us to
benefit from
the DMS efficiency within each node,
\julien{by leveraging} 
the embarrassingly parallel
computation of the discrete gradient which is the first, most time-consuming, DMS step.
The subsequent DMS steps are however less suitable for efficient
distributed computations\majorRevision{, in particular the pairing step\pierreRevision{s} to extract $\diagram_0$, 
$\diagram_1$ and $\diagram_2$}. Like most TDA algorithms, they indeed provide a
global view point on the data, which
%\majorRevision{. This} %% Pierre : je trouve que la phrase est trop découpée                                
requires multiple 
global data traversals with little computation\majorRevision{. This} combination of factors
challenges scalability due to the additional communications,
synchronizations and computations required to ensure the correctness of 
the algorithm. \majorRevision{For example, computing one pair of the diagram 
may require to access the data of multiple processes. This creates significant 
communication overhead. Furthermore, the pairing should happen in a specific order,
creating strong data dependencies across all processes.}

These subsequent steps require thus an important redesign for DDMS. 
Besides, DDMS relies on a hybrid MPI+thread design %algorithm for persistence diagram computation from scalar data
combining the advantages of both shared and distributed memories: (i)
larger total memory and 
(ii) reduced communication and synchronization overheads, along with improved intra-node 
dynamic load balancing. \majorRevision{
Even on a single node, DDMS achieves average performance comparable to DMS while 
requiring less memory, thereby rendering \pierreRevision{DDMS}
%the algorithm %% Pierre : il me semble que ce sont tes structures de
%données (plutot que l'algorithme) qui offrent un gain en mémoire 
advantageous for single-node 
execution as well.    
In this work, though the original DMS was designed 
for both structured and unstructured grids, we focus on structured grids.}

\subsection*{Contributions}
\label{sec_contributions}

This paper makes the following new contributions:
\begin{enumerate}
    \item \emph{ A hybrid distributed, shared-memory algorithm for 
    the computation of persistence diagrams for 1D, 2D and 3D data:} 
    Our work extends the fastest shared-memory approach (DMS) to the distributed setting, 
    enabling an efficient computation of persistence diagrams for scalar fields of 
    significant sizes. This is achieved thanks to the following novel procedures:
    \begin{enumerate}
        \item A novel self-correcting distributed algorithm for computing the $\diagram_0$ and 
        $\diagram_2$ pairings, extending the related procedure of 
        DMS \cite{guillou_tech22} to the distributed setting. 
        \majorRevision{The new self-correction mechanism removes data 
dependencies, allows for parallel 
        computation and reduces the need for synchronization.}
        \item A novel procedure for computing $\diagram_1$, based on an efficient 
        extension of the "PairCriticalSimplices" procedure \cite{guillou_tech22}.
        \majorRevision{This new procedure introduces specific data structures 
        and computation tokens to \pierreRevision{handle} %compensate
        %for
        the distribution of data 
        across multiple processes and ensure\julienRevision{s} the correctness 
of the algorithm.} 
        \majorRevision{A dedicated communication thread is also introduced to \pierreRevision{overlap communication and
        computation, to} reduce the idle time due to synchronizations 
        and \pierreRevision{to} improve 
        communication reactivity.}
%         the reactivity of communications.}
    \end{enumerate}
    The algorithm is output sensitive and provides substantial gains over the original 
    DMS approach as well as \majorRevision{\emph{DIPHA}}, the reference method for computing 
persistence 
    diagrams in a distributed setting.
    \item \emph{An open-source implementation:} For reproduction purposes, we provide a 
    C++ implementation of our approach \pierre{using MPI+OpenMP and} based
    on the Topology ToolKit (TTK)
    \cite{ttk19, ttk17, leguillou_tech24} %, which is
    integrated in ParaView.
    \item \emph{A reproducible example}: We provide a reference Python
    script for computing the persistence diagram with a dataset size that 
    can be adjusted to fit the capacities of any system (publicly available at: 
    \url{https://github.com/eve-le-guillou/DDMS-example}). \majorRevision{The
      larger example on the Turbulent Channel Flow
      \pierreRevision{(\autoref{fig_dns})} can be produced using the
      pipeline available in this \pierreRevision{script}. %% example. %% Pierre : pour éviter répétition sur "example" 
    } 
\end{enumerate}

The rest of the paper is organized as follows. In \autoref{sec_background}, we provide the necessary theoretical
foundations, %  needed in this work,
as well as a comparison to other existing methods.
The original DMS algorithm is summarized in \autoref{sec_originalDMS}.
\autoref{sec_overview} offers a high-level overview of DDMS. 
In \autoref{sec_extr-sad}, we provide a detailed explanation of the
distributed computation
\julien{of $\diagram_{0}$ and $\diagram_{2}$, including our novel
self-correcting procedure at the core of this computation.}
% %
% of stable and unstable sets
% \pierre{[P2E : enlever
%     mention ``stable and unstable sets'']}
%     and
% of our self-correcting procedure for the construction of $\diagram_{0}$ and $\diagram_{2}$.
In \autoref{sec_sad-sad}, we present the distributed computation of
$\diagram_{1}$ which has been designed by revisiting the DMS procedure ”PairCriticalSimplices".
Our experiments are presented in \autoref{sec_results}, both in strong and weak scaling.
We also demonstrate significant gain over \majorRevision{\emph{DIPHA}} \cite{dipha} 
(\autoref{sec_perfComparison}),
\julien{to our knowledge}
the only
publicly available implementation \julien{for} this problem in a
distributed context. Finally, we illustrate the
new distributed capabilities of DDMS by computing the persistence diagram (\autoref{sec_bigExample}) of 
a
\julien{3D scalar field}
% dataset
of 6 billion vertices distributed on 16 nodes of 32 cores each.

\section{Background}
\label{sec_background}

This section presents several theoretical elements of computational 
topology. For a more comprehensive introduction to this topic, we refer the 
reader to the reference textbook \cite{edelsbrunner09}.

\subsection{Input Data}
\label{sec_inputData}
The input domain  $\domain$ is a $d$-dimensional simplicial complex, with $d \leqslant 3$ in our applications.
A simplicial complex can be intuitively defined as a
\julien{discretization of a topological space}
% locally smooth topological space
% discretized
using small building blocks of different dimensions (a $0$-simplex is a vertex,
a $1$-simplex an edge, a $2$-simplex a triangle and a $3$-simplex a tetrahedron). A 
\emph{face} $\tau$ of a \pierre{$d$-simplex} $\sigma$ is a simplex defined by a non-empty
subset of the $d + 1$ vertices of $\sigma$. For example, 
the faces of a tetrahedron are its four triangles, its six edges and four vertices.
Inversely, $\sigma$ is called the co-face of $\tau$. 
On the domain $\domain$ is defined a piecewise linear (PL) scalar field $f : \domain \rightarrow \mathbb{R}$.
The input field $f$ is our subject of interest and is provided on the vertices
of $\domain$.
% \julien{(it is extended on the simplices of highter )}.
% It is interpolated on the simplices of higher dimension.
% using the Freudenthal
% triangulation (for regular grids).
$f$ is assumed to be injective on the vertices\julienRevision{. This is
enforced in practice by}
% which
% is
% achieved
% by
\julienRevision{a procedure inspired from  \emph{Simulation of Simplicity}
\cite{edelsbrunner90}\majorRevision{. It} consists in}
\julienRevision{pre-sorting the vertices by increasing $f$ values
(breaking ties according to the vertex offsets in memory) and by finally
considering as input data the vertex
orders
% positions
in this global data sort.}

% \majorRevision{considering an additional injective offset field, usually
% equal to the index of the vertices in the triangulation. The offset is
% used to disambiguate vertices of same scalar values. For efficiency, a global
% vertex order is generated once at the beginning of the computation and then
% the value of a vertex is substituted by } its
% position in the \julien{global} vertex order (by increasing $f$ values), a
% practice inspired
% from \emph{Simulation of Simplicity} \cite{edelsbrunner90}.

\begin{figure}
	\centering
	\includegraphics[width=\linewidth]{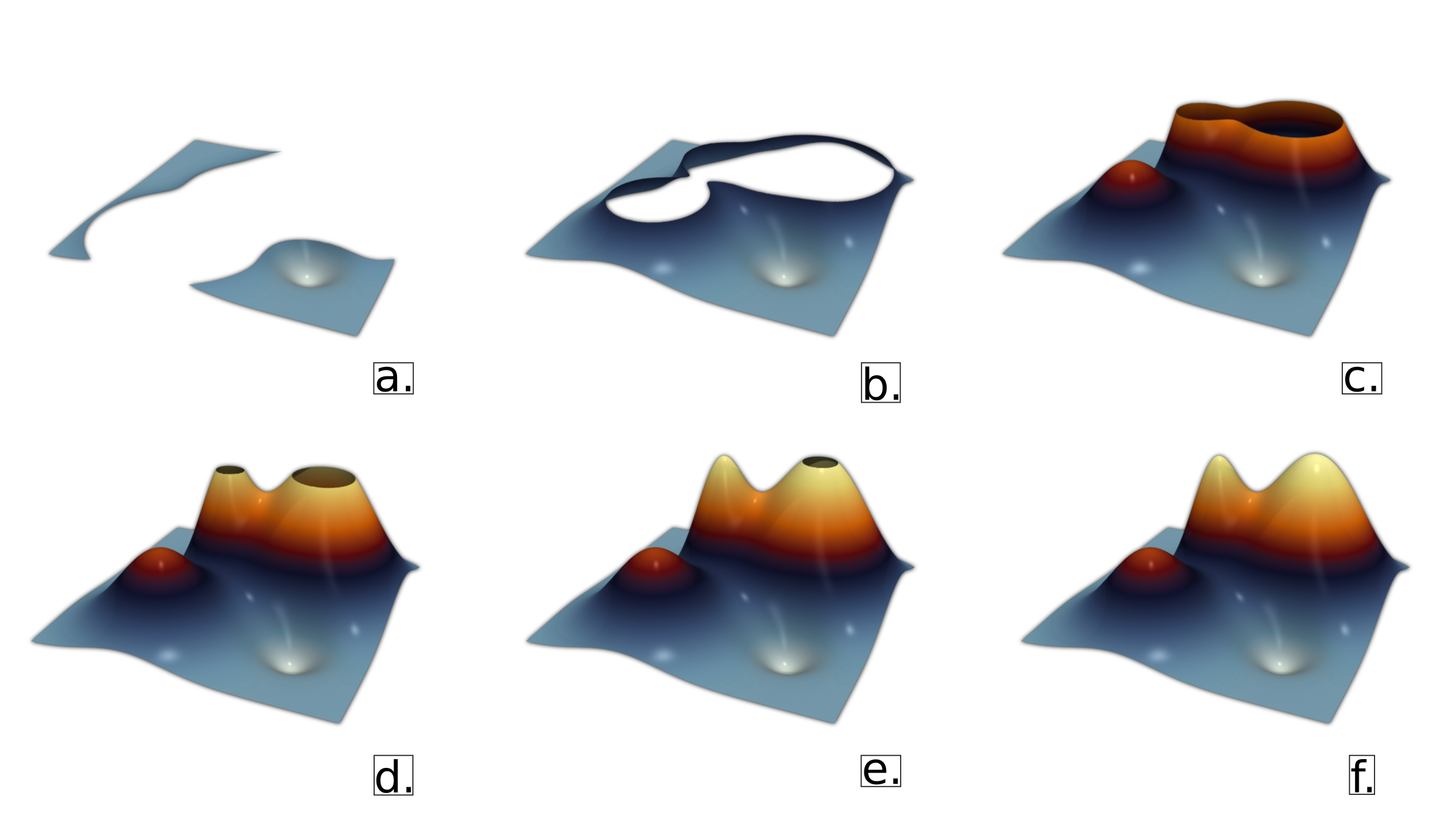}
	\caption{\julien{Filtration example for} a PL scalar field representing
the elevation on a terrain.
  \julien{As simplices are added by increasing scalar value,}
% 	As the next simplex of lowest scalar value is added to the filtration,
	the topology of the
	domain changes: in (a), the terrain is composed of two connected components.
	In (b), only one connected component remains, but a handle has appeared.
  In (c), the smaller hill at the front has been filled and there is still a handle.
  In (d), there are two handles. In (e), only one handle is left. 
  Finally, in (f), the whole domain has been swept \julien{(one connected
component)}.}
	\label{fig_filtration}
\end{figure}

% As $f$ is injective,
% a global order of all simplices of $\domain$ can be defined.
\julien{When the input is provided on a regular grid, it is implicitly
decomposed into a simplicial complex by following the Freudenthal triangulation
\cite{freudenthal42}.}
\majorRevision{This method will subdivide cubical cells into several simplices. 
For example, in 2D, a square is divided in two triangles (as can
\julienRevision{be} seen in \autoref{fig_vpaths}). 
This design choice is central to TTK and allows a genericity of support which is also
characterized by implementation simplicity. Furthermore, this particular triangulation yields a
natural interpolation within the input grid which coincides with the critical simplices identified by
discrete Morse theory.}
The comparison of two simplices is made by listing the vertex orders of both simplices by 
decreasing values, and by comparing the resulting lists with \emph{lexicographic} comparison.
\majorRevision{This comparison provides a strict global order over all simplices of $\domain$.}
Let $\domain^i$ be the union of the first $i$ simplices of $\domain$ of this global order.
The \emph{lexicographic filtration} of $\domain$ is the nested sequence of simplicial 
complexes $\emptyset \ne \domain^0 \subset \domain^1 \subset \dots \subset \domain^n = \domain$ 
with $n$ the number of simplices of $\domain$. \autoref{fig_filtration} shows an example of 
filtration. \majorRevision{For a more
\julienRevision{detailed}
% precise
definition of the lexicographic
filtration, we refer the reader to \cite{guillou_tech22}.}
The persistence diagram (defined in \autoref{sec_persistenceDiagram}) captures changes in the topology of the sub-complexes 
as the filtration evolves, making the filtration central to its construction.

\subsection{Distributed Model}
\label{sec_distributedModel}

The domain $\domain$ is divided between $\numberProcesses$ processes in
\julien{$\numberProcesses$}
disjoint blocks of data.
$f$ is defined on those blocks.
Each process $p \in \{0, \dots, \numberProcesses-1\}$ is associated with a local block 
$f_p : \domain_p \rightarrow \mathbb{R}$, such that:
\begin{itemize}
 \item $\domain_p \subset \domain$: each block $\domain_p$ is a
 $d$-dimensional simplicial complex and a subset of $\domain$.
 \item  \julien{A simplex $\simplex$ is \emph{exclusively owned} by
the process with the lowest identifier $p$ whose block $\domain_p$ contains
$\simplex$.}
%  Any simplex $\simplex$ present in multiple blocks (e.g., at the boundary
% between adjacent blocks)
%  is said to be \emph{exclusively owned} by the process with the lowest
%  identifier.
  \item A simplex $\simplex \in \domain_p$ which is not exclusively owned by  the process $p$ is called a \emph{ghost simplex} (see also below).
  \item $\cup_{\domain_p} = \domain$: the union of the blocks is equal to
the input.
\end{itemize}
A layer of additional \emph{ghost simplices} is added to each
\julien{block}
$\domain_p$.
\julien{We note $\domain'_p$ the
$d$-dimensional simplicial complex obtained by considering a layer of
ghost simplices, i.e.,
by adding to $\domain_p$
the
$d$-simplices of $\domain$ which share a face with a $d$-simplex of
$\domain_p$,
along with all their
$d'$-dimensional faces (with $d' \in \{0, \dots, d-1\}$).
Overall, all the simplices added in this way to the block $\domain_p$ to form
the \emph{ghosted block} $\domain_p'$ are
referred to as
\emph{ghost simplices}.}
% Such additional ghost simplices are simplices
% inside the block of a process that are copies
% of the interfacing simplices of an adjacent block.
% A \emph{ghosted block} is noted $\domain'_p$. All the simplices added in this way to the
% block $\domain_p$ to form the \emph{ghosted block} $\domain_p'$ are
% referred to as \emph{ghost simplices}.
In practice, the partitioning of our domain and the generation of ghosts is performed using ParaView \cite{paraviewBook}, 
a well-known visualization software. This preprocessing step is
considered as already performed on our input data.

Our work is implemented in TTK and makes use of its features,
in particular, the triangulation. The triangulation is a data structure
that allows for efficient query execution and data traversal. It has been recently
ported to a distributed-memory setting \cite{leguillou_tech24}, which induces specific preconditioning steps
to instantiate the triangulation, such as the ghost layer generation or the computation 
of the local adjacency graph of $\domain_p$. This preconditioning overhead is negligible
compared to already existing distributed algorithms of TTK.

\subsection{Discrete Morse Theory}
\label{sec_dmt}

\begin{figure}
  \centering
  \includegraphics[width=\linewidth]{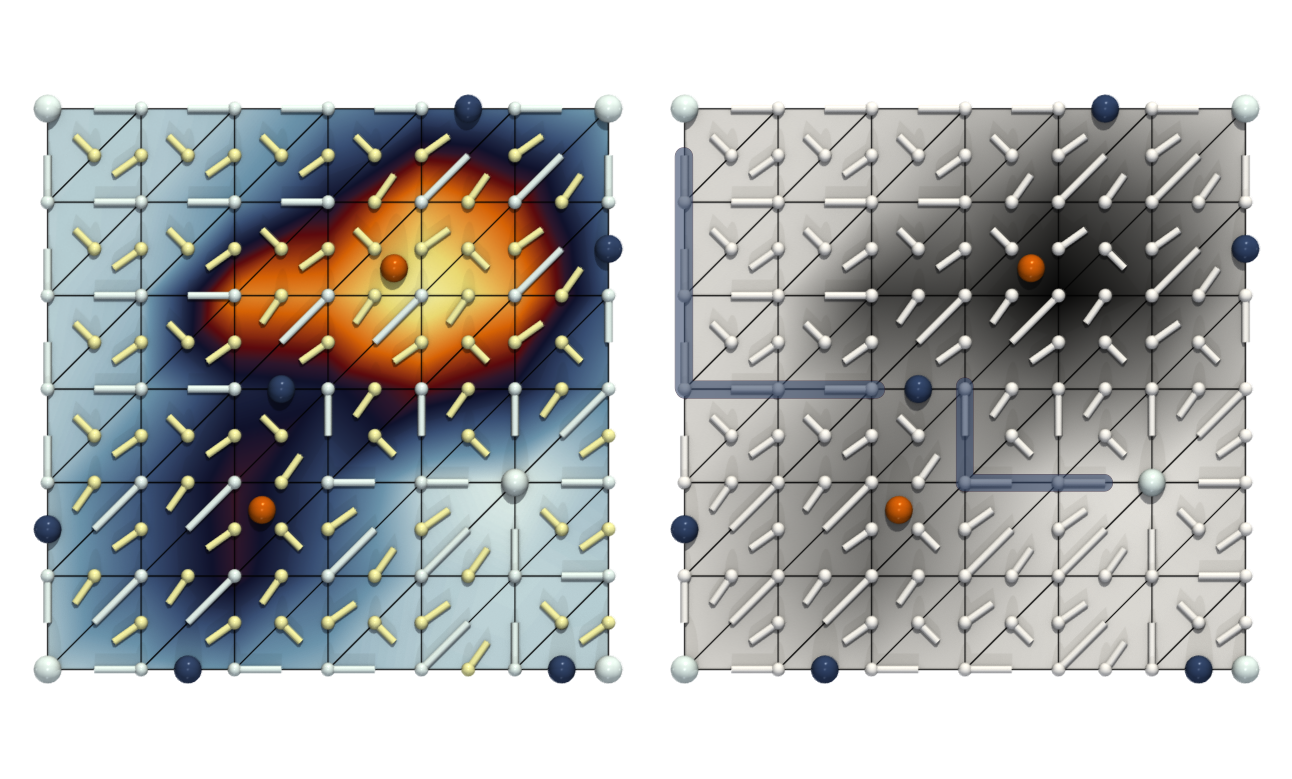}
  \caption{Example of discrete gradient field for a given scalar field (left).
  The larger spheres represent critical simplices. The light yellow arrows are edge-triangle 
  pairs\julien{. The} light blue arrows are vertex-edge pairs. Two examples
of v-paths are
  shown in transparent blue \julien{(right)}, going from a critical edge and
ending in critical vertices.
\julien{These form the unstable set of the critical edge.}}
%   These particular v-paths correspond to unstable sets.}
  \label{fig_vpaths}
\end{figure}

In this section, we introduce notions of \emph{Discrete Morse Theory} (DMT) \cite{forman98} 
which is central to DMS.
A \emph{discrete vector}  is a pair formed by a simplex $\simplex_i \in
\domain$ (of dimension $i$) and one of its co-facets $\simplex_{i+1}$ (i.e., one
of its co-faces of dimension $i+1$), noted $\{\simplex_i < \simplex_{i+1}\}$.
$\simplex_{i+1}$ is usually referred to as the head of the vector, while $\simplex_{i}$ is
its tail.
Examples of discrete vectors include a pair between a vertex and one of its incident edges, 
or a pair between an edge and a triangle containing it. A \emph{discrete vector field} on $\domain$ is then
defined as a collection $\discreteVectorField$ of pairs $\{\simplex_i <
\simplex_{i+1}\}$, such that each simplex of $\domain$ is involved in at most
one pair. A simplex $\simplex_i$ which is involved in no discrete vector
$\discreteVectorField$ is called a \emph{critical simplex}.

A \emph{v-path} is a sequence of discrete vectors
$\big\{\{\simplex^0_i < \simplex^0_{i+1}\}, \dots, \{\simplex^k_i <
\simplex^k_{i+1}\}\big\}$, such that the tails of two consecutive vectors are distinct and
the tail of a vector in the sequence is a face of the head of the previous
vector (see \autoref{fig_vpaths}, right).
The collection of all the v-paths terminating in a given critical simplex
$\sigma_i$ is called the \emph{discrete stable set} of $\sigma_i$.
Symmetrically, the collection of all the discrete v-paths
starting at a given critical simplex $\sigma_i$ is called the \emph{discrete
unstable set} of $\sigma_i$. Finally, a \emph{discrete gradient field} is
a discrete vector field such that all its possible \emph{v-paths} are loop-free
(see \autoref{fig_vpaths}, left).
Intuitively, this means that all critical simplices can be connected by following discrete 
vectors without any loops in the paths.
The dimension of the critical simplex corresponds to its index in the smooth
setting \cite{milnor63, morseQuote}. A critical $0$-simplex (or vertex) is
called a minimum, a $1$-simplex (or edge) a $1$-saddle, a $2$-simplex (or
triangle) a $2$-saddle and a $3$-simplex (or tetrahedron) a maximum.
% Critical
% simplices can be linked by \emph{v-paths}.
 
Several algorithms have been
proposed to compute such a discrete gradient field from an
input PL scalar field. We consider in this work the algorithm by Robins et al.
\cite{robins_pami11}, given its proximity to the PL setting: each
critical cell identified by this algorithm is guaranteed to be located in the
star of a PL critical vertex. In practice the computation is performed through 
inspection of the local neighborhood of each vertex,
which makes this step embarrassingly parallel. 

\subsection{Homology groups}

\begin{figure}
	\centering
	\includegraphics[width=0.7\linewidth]{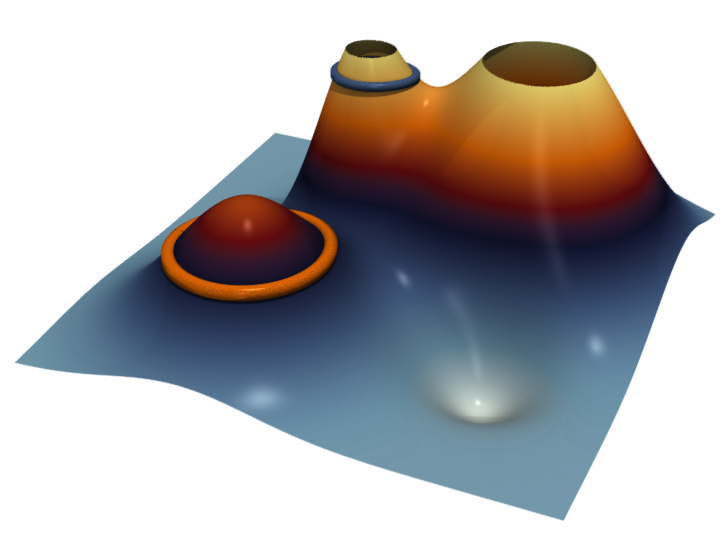}
	\caption{Examples of $1$-cycles on a sub-complex of a filtration.
	The PL scalar field represents the elevation on a terrain. Both the orange and the 
	blue cycles are $1$-cycles: the computation of their boundaries
results in
	summing all the vertices of each cycle twice (as all vertices are faces of exactly 
	two edges of the cycle), yielding a boundary equal to $0$ for each cycle.
	However, the two cycles are not homologous: it is not possible 
	to go from one to the other by
% 	simply
	adding a $p$-boundary. The blue cycle
  can be chosen as the
%   generator
  \julien{representative}
  of its class, representing the hole at the top of
  the hill it surrounds.}
	\label{fig_cycles}
\end{figure}

In this section, we introduce key concepts of topology relative to homology groups 
used in the original DMS algorithm. Intuitively, homology relies on the concept of
cycles, used to detect "holes" in the data. In 3D, "holes" correspond to connected components,
handles and voids. Cycles are a sum of simplices, or a collection of simplices, that can "go around" the domain and end 
where they started. \autoref{fig_cycles} provides a visual example of these notions. 
Here are a series of definitions to formalize these concepts. 
A $p$-chain $c$ is a formal sum of modulo 2 coefficients of $p$-simplices $\sigma_i$ 
of $\domain$: $\sum \alpha_i\sigma_i$ with $\alpha_i \in \{0,1\}$. 
Two $p$-chains can be summed together component-wise to form a new $p$-chain.
A $p$-chain can be modeled with a bit mask where a simplex is present 
if it has been added an odd-number of times.
For example, adding the two $0$-chains $a = v_0 + v_1$ and $b = v_1 + v_2$ will yield 
$ a + b = v_0 + v_2$. 
The boundary of a $p$-simplex $\sigma_i$ is noted
$\partial \sigma_i$ and is defined as the sum of its faces of dimension $(p-1)$.
The boundary of a $p$-chain $c$ is the sum of the boundaries of its simplices: 
$\partial c = \sum \alpha_i \partial \sigma_i$. A $p$-cycle $c$ is a $p$-chain such 
that $\partial c = 0$. The group of all $p$-cycles is noted $\mathcal{Z}_p(\domain)$.

Cycles are not sufficient to detect the "holes" of a domain. One must look into the relationship 
between cycles and $p$-boundaries. A $p$-boundary is a $p$-chain that is itself the boundary of
a $(p+1)$-chain. The group of p-boundaries of a simplicial complex $\domain$ is the group 
of all $p$-boundaries of $\domain$, noted $\mathcal{B}_p(\domain)$. 
The $p^{th}$ \emph{homology group} of a simplicial complex $\domain$ is
the quotient group of its $p$-cycles modulo its $p$-boundaries: 
$\mathcal{H}_p(\domain) = \mathcal{Z}_p(\domain)/\mathcal{B}_p(\domain)$. 
Specifically, two $p$-cycles $c$ and $c'$ are called \emph{homologous} if one can be transformed
into the other by adding the boundary of a $(p+1)$-chain $c''$: $c = c' + \partial c''$. 
The set of all cycles that are homologous defines a \emph{homology class}. Any cycle 
of the class can be chosen as a representative.
Intuitively, a homology class gathers all the cycles \julien{surrounding}
the same \julien{set of ``holes''}.
\julien{The rank of the $p^{th}$ homology group $\mathcal{H}_p(\domain)$, also
known as the  $p^{th}$ Betti number, is the number of linearly independent
classes of $\mathcal{H}_p(\domain)$ (called generators).}
% Such a representative is called the
% \emph{generator} of the class.
Detecting \julien{the $p$-dimensional independent} \emph{``holes''} in a domain
therefore means \julien{extracting generators for the $p^{th}$ homology group.}
% generators
% for all the homology classes of
% a homology group.

\subsection{Persistence diagrams}
\label{sec_persistenceDiagram}

\begin{figure}
  \centering
  \includegraphics[width=\linewidth]{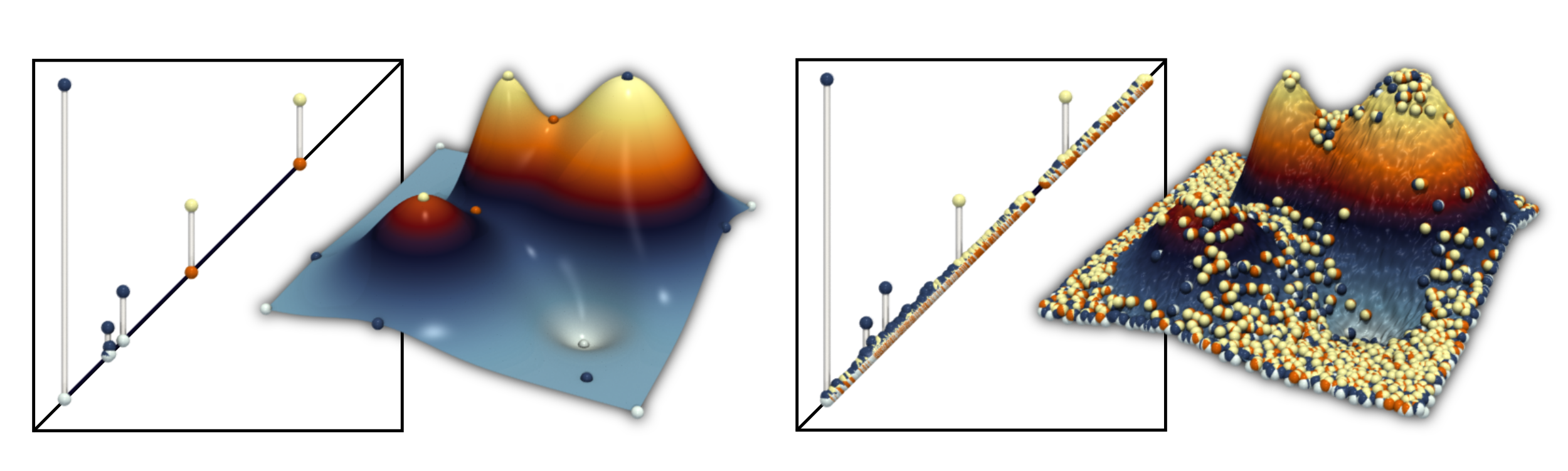}
  \caption{Persistence diagram for a clean (left) and noisy (right) dataset. Critical 
  simplices are denoted by spheres (light blue: minima, light yellow: maxima, others: saddles). The persistence of 
  each pair is the height of the bar. $\diagram_{0}$ corresponds to minimum-saddle pairs 
  and $\diagram_{2}$ corresponds to saddle-maximum pairs. 
  Salient features (long pairs) can easily be distinguished from the noise (short pairs).}
  \label{fig_persistenceDiagram}
\end{figure}

As we browse %go through
the filtration, the topology of the domain evolves as shown in \autoref{fig_filtration}.
The persistence diagram of dimension \julien{$p$}, noted \julien{$\diagram_p$},
tracks the evolution of \julien{$p^{th}$} homology generators during a
filtration.
In particular, $\diagram_0$ tracks the evolution of 
connected components, $\diagram_1$ of handles and $\diagram_2$ of voids.
\julien{Specifically, at the step $i$ of the filtration, a new generator
$\gamma$ can appear upon the insertion of a $p$-simplex $\simplex_i$.
For instance, a new $1$-cycle $\gamma$ can appear by adding an edge which
connects two vertices of $\domain_i$ belonging to the same connected component.
In that
case, $\gamma$ is said to be \emph{born} at step $i$.
Further in the filtration, at step $j$, $\gamma$ can disappear upon the
insertion of a $(p+1)$-simplex $\simplex_j$ which completes the
filling of the \emph{hole} created at step $i$ (i.e., in $\domain^j$, $\partial
\simplex_j$ is \emph{homologous} to the cycle created by $\simplex_i$ in
$\domain^i$).
For instance, a $1$-cycle $\gamma$ made of three edges can be filled upon the
introduction of the triangle containing these three edges.
Then $\gamma$ is said to \pierre{\emph{die}} at step $j$. The simplices
$\simplex_i$ and $\simplex_j$ form a $p$-dimensional \emph{persistence pair}.
As shown in \autoref{fig_persistenceDiagram}, each $p$-dimensional persistence
pair is embedded in the persistence diagram $\diagram_p$, as a point in the
so-called birth/death plane, at coordinates $\bigl(f(\simplex_i),
f(\simplex_j)\bigr)$, where $f(\simplex)$ is the highest data value among the
vertices of $\simplex$. The \emph{persistence} of a pair can be read as the
height of the point to the diagonal (i.e., $f(\simplex_j) - f(\simplex_i)$).}
%
% Each persistence
% pair is embedded at a single opint in the
%
%
% this generator is said to \emph{die}
% upon the insertion of a critical $p+1$-simplex $\sigma_j$ which completes the
% filling of the \emph{hole} created at step $i$ (i.e., in $\domain^j$, $\partial
% \sigma_j$ is \emph{homologous} to the cycle created by $\sigma_i$ in
% $\domain^i$).}
%
% Each homology generator of
% % $\domain^p$
% \julien{$\domain$}
% can be associated with a unique pair of critical
% simplices $(c, c')$, corresponding to its birth and death, the simplices at which the feature appears and disappears.
% The birth corresponds to the creation of the homology class and its
% According to the Elder rule \cite{edelsbrunner09}, critical simplices can be arranged in pairs
% such that each simplex appears in only one pair, with $f(c) < f(c')$, $c$ being a $p$-simplex
% and $c'$ a $(p+1)$-simplex. For example, if two connected components,
% born respectively in critical vertices $v$ and $v'$, with $f(v) < f(v')$, meet at the critical edge $e$,
% the vertex of higher value, here $v'$, dies, or stops existing as a separate connected component,
% in favor of $v$, the vertex of lower value. The pair $(v', e)$ is then added to
% the persistence diagram.
%
% As shown in \autoref{fig_persistenceDiagram}, each pair of the diagram is embedded in a single point in the birth/death 2D plane at the
% coordinates $\bigl(f(c), f(c')\bigr)$. The persistence of a pair can be read as
% the height of the point from the diagonal (i.e. $f(c) - f(c')$).
Salient features are characterized by a high persistence,
% and correspond to
\julien{with}
the points located far from the diagonal. Conversely, noise in the data
produces points in the plane close to the diagonal.

\subsection{The original Discrete Morse Sandwich algorithm}
\label{sec_originalDMS}

Here is an overview of the original Discrete Morse Sandwich algorithm. For a more detailed description of 
the algorithm, we refer the reader to its introduction paper \cite{guillou_tech22}.
We consider here 3D datasets, where  $\diagram_0$, $\diagram_1$ and
$\diagram_2$  have to be computed. 
%%For 2D datasets, only the diagrams $\diagram_0$ and $\diagram_1$ are computed as follows.
First, the discrete gradient is computed in parallel using multi-threading. The critical simplices are
then deduced from 
the gradient. This step is called the \emph{zero-persistence skip}: \julien{each
remaining, non-critical simplex forms a zero-persistence pair with the other
simplex involved in its discrete vector (\autoref{sec_dmt})}. The rest of the
algorithm will
focus on  pairing the
obtained critical simplices, following a stratification strategy where
$\diagram_0$ and $\diagram_2$ \julien{(special cases)} are computed
before $\diagram_1$.

\begin{figure}
  \centering
  \includegraphics[width=\linewidth]{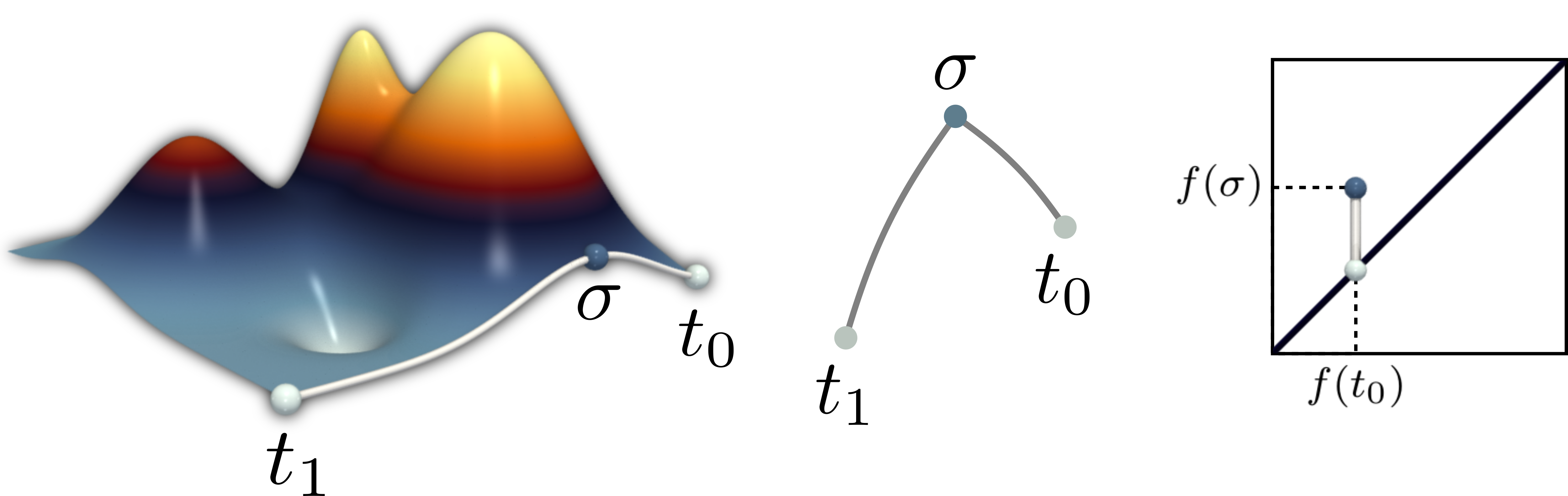}
  \caption{Overview of the DMS algorithm for the computation of $\diagram_0$. First,
  the unstable set is computed from the vertices of the critical $1$-simplex $\sigma$
  by tracing v-paths (white curves, left). The set is then collapsed into
  \julien{an \emph{extremum graph}}
%   the
% graph
%   $\graph{0}(f)$
  \julien{$\graph{0}$}
  (middle).
  Each critical simplex is represented by a node in the graph, with the arcs representing the v-paths.
  Finally, the graph is processed with a Union-Find structure to produce the persistence 
  pair in $\diagram_0$ (right).}
  \label{fig_D0}
\end{figure}

The diagram $\diagram_0$ is computed first. An overview of this algorithm is shown 
in \autoref{fig_D0}. 
We start by building
\julien{an extremum} graph \julien{noted} $\graph{0}$ by following, for the two
vertices $v_0$ and
$v_1$ of a critical edge $\sigma$, the gradient until a critical $0$-simplex (or extremum) 
is reached ($t_0$ and $t_1$). Each critical edge is processed in parallel (multi-threading). 
\julien{If $t_0$ and $t_1$ are distinct, the}
% The
triplet $(\sigma, t_0, t_1)$ represents new elements of the graph
$\graph{0}$,
adding the three nodes $\sigma$, $t_0$ and $t_1$ (one per element of the triplet) and two arcs representing 
the v-paths between $\sigma$ and the two extrema. 
When all critical $1$-simplices are processed, $\graph{0}$ is complete. $\diagram_0$ is then
computed by visiting the edges of $\graph{0}$ in increasing order following 
the
\julien{PairExtremaSaddles}
% PairExtremumSaddle
algorithm presented in \autoref{alg_pairExtremumSaddle}.
This step is intrinsically
sequential, and relies on a Union-Find data structure for each 
node of $\graph{0}$. A Union-Find efficiently models connectivity in data through two 
primitives: \emph{find()}, that returns the representative of
\julien{the connected component containing}
the node,
and \emph{union()}, that merges together two
% nodes
\julien{components}
by unifying their representative\julien{s}.
Initially, each $0$-simplex is its own representative. For each triplet $(\sigma, t_0, t_1)$ of $\graph{0}$, 
the following procedure is applied: the representatives $r_0$ and
$r_1$ of $t_0$ and $t_1$ are retrieved (using  \emph{find()}). %computed. 
$r_0$ is ensured to be the highest representative (l. 5-6 in \autoref{alg_pairExtremumSaddle}), by
swapping if necessary its value with $r_1$. % in case it is not.
The highest representative, $r_0$, is then paired with $\sigma$ and is assigned $r_1$ as representative.
In \autoref{fig_D0}, this step creates the pairing $(\sigma, t_0)$, with $t_1$ the new representative of $t_0$.
To speed up the computation, the triplets of $\graph{0}$ are collapsed as they are visited. This means that 
the representative of $t_0$ is also set to $r_1$ (l. 12), which  %This
is equivalent to a \emph{path compression} in a Union-Find data structure. 
This concludes the computation of $\diagram_0$.

The diagram %$\diagram_{1}$, respectively
$\diagram_{2}$ %for 3D datasets, 
is computed similarly on critical %$1$- and $2$-simplices %(respectively
$2$- and $3$-simplices % for 3D datasets, ),
using a dual discrete gradient field, obtained by reversing every discrete vector of the discrete gradient 
on the domain's dual complex (see \cite{guillou_tech22} for more details).

\begin{algorithm}
  \caption{\julien{PairExtremaSaddles}}\label{alg_pairExtremumSaddle}
  \begin{algorithmic}[1]
      \REQUIRE An ordered set $C_{1}$ of triplets $(\sigma_j, t_0, t_1)$ of \pierre{$\mathbf{\graph{0}}$}
      \ENSURE \majorRevision{The persistence diagram $\diagram_{0}$, a collection of pairs ($\sigma, t$)}
      \FOR[Process the $1$-simplex $\sigma_j$]{$j \in C_{1}$}
        \STATE $r_0 \gets findRepresentative(t_0)$
        \STATE $r_1 \gets findRepresentative(t_1)$
        \IF{$r_0 \ne r_1$}
          \IF{$r_0 < r_1$}
            \STATE swap$(r_0, r_1)$
          \ENDIF
          \STATE addPair$(\sigma_j, r_0)$
          \STATE $Representative[r_0] \gets r_1$
          \STATE $Representative[t_0] \gets r_1$    
          \hfill\COMMENT{The arc is collapsed}        
        \ENDIF
      \ENDFOR
      \end{algorithmic}
  \end{algorithm}
\begin{figure}
  \centering
  \includegraphics[width=\linewidth]{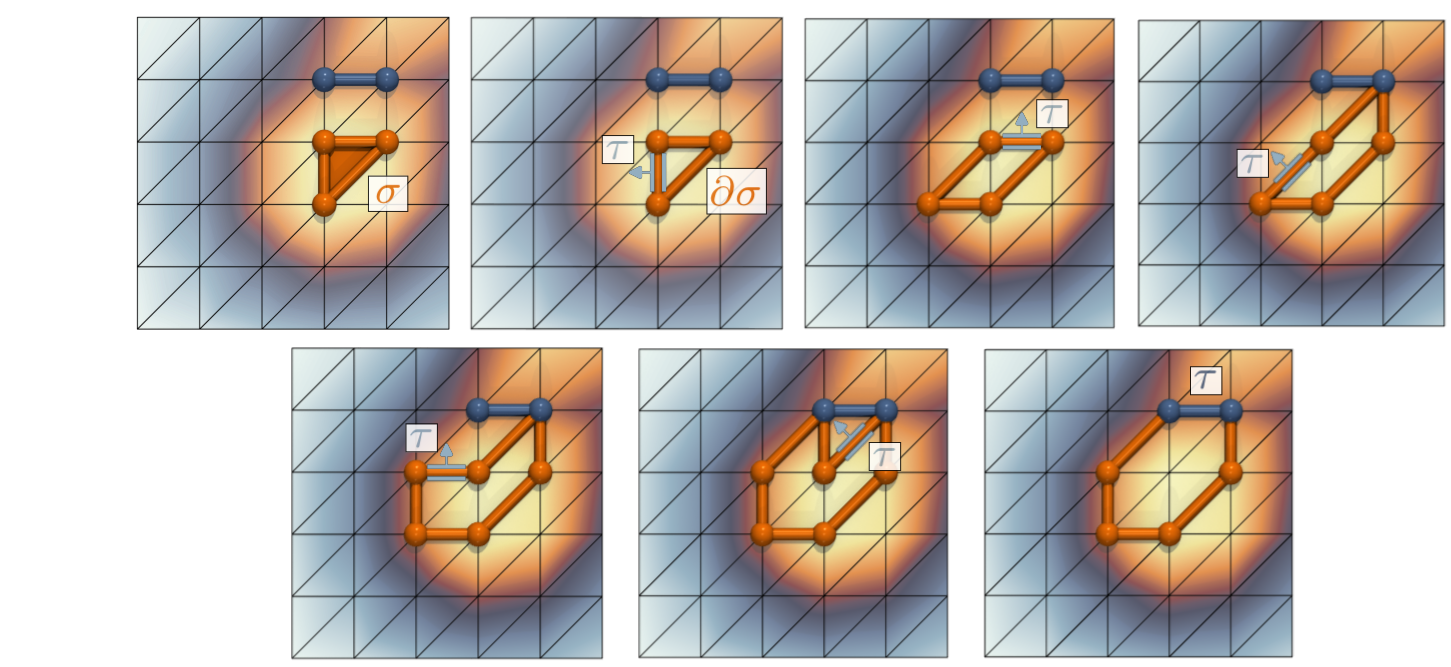}
  \caption{A homologous propagation given a simplex $\sigma$. The boundary of 
  $\sigma$, $\partial \sigma$, is iteratively expanded by selecting its highest $1$-simplex $\tau$ and 
  adding to the boundary of $\sigma$ the boundary of the $2$-chain associated to $\tau$.
  The boundary is expanded until reaching a critical $1$-simplex (here in dark blue).
  The pair $(\sigma, \tau)$ is then added to $\diagram_1$.}
  \label{fig_homologousProp}
\end{figure}

For 3D datasets, %% $\diagram_0$ and $\diagram_2$ are computed using
%% the previous procedure, and
$\diagram_1$ is computed last.
The first step of the construction of $\diagram_1$ consists in
restricting its input to the unpaired critical $1$- and $2$-simplices:
the critical $1$- and $2$-simplices already paired in $\diagram_0$ or $\diagram_2$ 
are hence not considered here. This stratification strategy greatly reduces the number
of input simplices for $\diagram_1$.
We then apply \autoref{alg_pairCriticalSimplices} using the
PairCriticalSimplex procedure defined in \autoref{alg_pairCriticalSimplex}.
For each unpaired critical $2$-simplex $\sigma$, a \emph{homologous propagation} (See \autoref{fig_homologousProp}) is performed. 
It expands a \julien{boundary,} initially equal to $\partial \sigma$, by selecting the highest
$1$-simplex $\tau$ of the \julien{current boundary}
and adding \julien{to it}
% to the boundary of $\sigma$
the boundary of the $2$-chain associated
to $\tau$.
For simplicity, we will refer in the remainder of this paper to
\julien{the boundary}
% the $2$-chain
initiated \julien{this way} in $\sigma$ as the \emph{boundary of $\sigma$}.
The propagation stops
at the first unpaired $1$-simplex. The pair $(\tau, \sigma)$ can then
be created. $\sigma$ is the death and $\tau$ the birth of the pair.
\julien{Note that the}
% The
propagation
is expended in reverse relative to the filtration order\julien{: given a
triangle $\sigma$, this process identifies the edge $\tau$ which created the
latest $1$-cycle $\gamma$ in the filtration which is \emph{homologous} to
$\partial \sigma$ (\autoref{sec_persistenceDiagram}).}
%% For now, we have assumed in our explanation that the input set of simplices is ordered.
All this assumes that the input set of simplices is ordered. 
However, as introduced by 
Nigmetov et al. in \cite{Nigmetov20}, it is possible to process the
simplices in a random manner, \pierre{hence in parallel (multi-threading)}.
An extra case has then to be considered: % needs to be added:
when performing the homologous propagation for a simplex $\sigma$,
it is possible to reach a $1$-simplex $\tau$ that has already been paired to a $2$-simplex $\sigma_{\tau}$
through homologous propagation. In that case, there are two possibilities: 
either (i) $\sigma_{\tau}$ is lower than $\sigma$ and the propagation carries on by merging 
$\partial \sigma_{\tau}$ with $\partial \sigma$, or (ii) $\sigma_{\tau}$ is higher than 
$\sigma$. In that case, the pair $(\tau, \sigma_{\tau})$ is removed, $(\tau, \sigma)$ is 
added and the homologous propagation of $\sigma_{\tau}$ is resumed.
%Additional
Compare-And-Swap operations are used %implemented to allow
for thread-safe memory accesses %for shared-memory  parallelism
as described by Nigmetov et al.\cite{Nigmetov20}.
%% The homologous propagation is described in \autoref{alg_pairCriticalSimplex}. 
%%
%%The routine is applied  in parallel (using multi-threading) on all unpaired 
%%$2$-simplices as describe in \autoref{alg_pairCriticalSimplices}.
$\diagram_1$ is finally created from the temporary pairs once all $2$-simplices have been completely processed.

  \begin{algorithm}
    \caption{PairCriticalSimplices}\label{alg_pairCriticalSimplices}
    \begin{algorithmic}[1]
        \REQUIRE Set $C_{2}$ of \julien{unpaired} critical $2$-simplices \majorRevision{$\sigma$}
        \ENSURE \majorRevision{The persistence diagram $\diagram_{1}$, a collection of pairs ($Pair(\sigma), \sigma$)}
        \FOR{$j \in C_{2}$ {\it in parallel (multi-threading)}}
          \STATE PairCriticalSimplex$(\sigma_j)$
        \ENDFOR
        \FOR{$j \in C_{2}$}
          \STATE $\diagram_{1} \gets \diagram_{1} \cup (Pair(\sigma_j), \sigma_j)$  
        \ENDFOR        
        \end{algorithmic}
    \end{algorithm}

\begin{algorithm}
  \caption{PairCriticalSimplex \emph{(homologous propagation)}}\label{alg_pairCriticalSimplex}
  \begin{algorithmic}[1]
      \REQUIRE An unpaired critical $2$-simplex $\sigma$
      \ENSURE A temporary pair of
%       $D_{1}(f)$
      \julien{$\diagram_{1}$}
      \IF{$\boundary{\sigma} == 0$}
      \STATE $\boundary{\sigma} \gets \partial \sigma$
      \ENDIF
      %\STATE \COMMENT{Homologous propagation of $\partial \sigma$}
      \WHILE{$\boundary{\sigma} \ne 0$}
        \STATE $\tau \gets$ max$(\boundary{\sigma})$
        \IF[Expand boundary]{$\tau$ is not a critical simplex}
          \STATE $\boundary{\sigma} \gets $
          \STATE \hskip\algorithmicindent $ \boundary{\sigma} + \boundary{Pair(\tau)}$
        \ELSE[$\tau$ is critical]
          \IF[$\tau$ is unpaired ]{$Pair(\tau) == \emptyset$}
            \STATE addPair$(\sigma, \tau)$         
            \STATE \textbf{break}
          \ELSE[$\tau$ has already been paired to $\sigma_{\tau}$]
            \STATE $\sigma_{\tau} \gets Pair(\tau)$
            \IF[Merge the boundaries]{$\sigma_{\tau} < \sigma$}
              \STATE $\boundary{\sigma} \gets $
              \STATE \hskip\algorithmicindent $ \boundary{\sigma} + \boundary{\sigma_{\tau}}$
            \ELSE[$\sigma$ is older and the true death of $\tau$]
            \STATE addPair$(\sigma, \tau)$
            \STATE $Pair(\sigma_{\tau}) \gets \emptyset $
            \STATE PairCriticalSimplex$(\sigma_{\tau})$ \COMMENT{Resume for $\sigma_{\tau}$}
            \ENDIF
          \ENDIF
        \ENDIF
      \ENDWHILE
      \end{algorithmic}
  \end{algorithm}

Regarding 2D datasets, only the diagrams
$\diagram_0$ and $\diagram_1$ have to be computed. $\diagram_1$ is
then computed in 2D on critical $1$- and $2$-simplices like $\diagram_2$ in 3D.

\subsection{Related work}

\textit{Persistent homology:} Multiple research groups independently introduced persistent homology
\cite{barannikovFramedMorseComplex1994, edelsbrunner02, frosini99, robins99}. 
In numerous data analysis applications, topological persistence rapidly emerged as a compelling measure 
of importance, helping in the identification of salient topological features within 
the data. The most common method for persistence homology computation relies on the reduction of the boundary matrix 
(that describes the facet/co-facet relations between the simplices 
of the input domain). DMS relies on a different strategy, based on
\julien{discrete}
Morse Theory \cite{forman98, perseusPaper, robins_pami11},
but there are several conceptual similarities
to existing documented accelerations.
%% P2E : "accelerations" non compris 
Bauer et al. \cite{ripser} introduced the idea 
of \emph{apparent pairs}, which is
% equivalent
\julien{similar}
to the \emph{zero-persistence skip} pre-computation step
of DMS. Furthermore, the stratification strategy
\julienRevision{used}
% employed
in DMS can be \majorRevision{linked with
the stratification used by Bauer et al. \cite{bauer2013clearcompresscomputingpersistent}}
through "\emph{Clearing}" and "\emph{Compression}"\majorRevision{. These operations} enables \majorRevision{one} to discard simplices
already involved in persistence pairs. Edelsbrunner et al. \cite{edelsbrunner02} observe that the persistence diagram
for dimension 0 can be computed through a Union-Find data structure. They also observe
that the 2-dimensional persistence diagram can be obtained, by symmetry, with 
a Union-Find structure. This was aggressively exploited by Guillou et
al. \cite{guillou_tech22} as they restricted the Union-Find 
% computation 
to stable and 
unstable sets of 1 and
% $(d-1)$
\julien{$2$}
saddles.

Other methods have investigated Morse Theory to accelerate the computation of persistent 
homology \cite{milnor63, morseQuote}, specifically in
a discrete setting \cite{forman98}. In particular, Robins et al \cite{robins_pami11} introduced 
the
% definition of the
% Discrete Gradient
\julien{discrete gradient}
employed in DMS and used it to accelerate
the computation of persistence. Other approaches improved 
on this idea \cite{GuntherRWH12, iuricich21, Wagner23} or extended it to a more
general setting not limited
to regular grids \cite{perseusPaper}.
\julien{For instance, Wagner introduced an out-of-core approach \cite{Wagner23}
capable of processing massive scalar datasets on commodity hardware, but at the
expense of significantly long computations
(typically hours of computation for
several billions of data points). In contrast, our approach targets
high-performance hardware in a distributed setting, with much faster
computations (\autoref{sec_bigExample}).
Also, note that this out-of-core approach \cite{Wagner23} is not included in the
performance benchmark by Guillou et al. \cite{guillou_tech22}, which was
published before\majorRevision{. However, }our preliminary experiments report a typical $\times 2$
speedup in favor of DMS \cite{guillou_tech22} on our hardware.
% (specified in \autoref{sec_results}).
Overall, in contrast to previous
approaches based on discrete Morse theory, DMS significantly accelerates the
process thanks to an aggressive stratification strategy, taking advantage of
the specificities of the diagrams $\diagram_0$ and $\diagram_2$ and avoiding
the computation of the full Morse complex.}
% DMS uses the idea of precomputation of the gradient to
% accelerate the construction of the persistence and further accelerates it by taking advantage
% of specificities of the $\diagram_0$ and
% % $\diagram_{d-1}$
% \julien{$\diagram_{2}$}
% diagrams.

In practice, there are numerous software packages 
available to produce persistence diagrams, such as \emph{PHAT} \cite{BauerKRW17}, 
\majorRevision{\majorRevision{\emph{DIPHA}}} \cite{dipha}, \emph{Gudhi} \cite{gudhi}, \emph{Ripser} 
\cite{ripser} 
or \emph{Eirene} \cite{eirene}.
Each implementation however focuses on particular data structures, such as
\julien{generic filtrations of}
cell complexes (for \emph{PHAT},
\majorRevision{\majorRevision{\emph{DIPHA}}} and \emph{Gudhi}) or Rips filtration
% and
\julien{of}
high-dimensional point
clouds (for \emph{Eirene} and \emph{Ripser}). Some of the listed software for persistence 
diagram computation are purely sequential (\emph{Eirene}, \emph{Ripser}), while 
others implement shared-memory parallelism (\emph{PHAT}, DMS). \majorRevision{Another 
implementation, \emph{Ripser++}, builds upon \emph{Ripser} to focus on
\pierreRevision{GPU acceleration,} %% Pierre : GPU ne s'utilise pas
%% trop au singulier : on  a GPU,
also using apparent pairs in a pre-computation step to speed up the overall algorithm.}
\majorRevision{C. T{\'o}th et al. \cite{toth2025user} also introduced GPU-accelerated computations of signature kernels, 
a machine learning tool relying on a dynamic sequence of persistence diagrams \cite{oberhauser2023}. Like \emph{Ripser++}, 
their implementation focused on Vietoris-Rips filtrations.}
\majorRevision{Otter et al. \cite{otterRoadmapComputationPersistent2017} provided in 2017 a 
benchmark focusing on execution speed and memory footprint on
\pierreRevision{one and multiple CPU cores}. %a single core and in shared-memory on a CPU. 
For Vietoris-Rips
\julienRevision{filtrations,}
% complexes,
\emph{Ripser} has been identified as the fastest 
% available
implementation,
whereas for \julienRevision{sub-level set filtrations on} cubical complexes,
\majorRevision{\emph{DIPHA}}
% has been
\julienRevision{was}
recognized
\julienRevision{then}
for its superior execution
speed.}
More recently, DMS
\julien{has been reported to}
% is proven to
be the fastest implementation using shared-memory parallelism, according to the benchmark
provided with its introductory paper \cite{guillou_tech22}.

\textit{Distributed-memory algorithms:} There are few existing
approaches relative to distributed-memory parallelism and persistence
diagrams. 
The most well-known method is \majorRevision{\emph{DIPHA}}, introduced by Bauer et al. in 
\cite{dipha}.
It computes the boundary matrix of the domain and partitions the matrix into blocks of contiguous columns. 
The boundary matrix represents the relations between the simplices and their faces. 
The matrix is then reduced using a variant of Gaussian elimination.
In particular, it is very similar to the spectral sequence algorithm for persistent 
homology \cite{edelsbrunner09}, with several adaptations to make it correct and 
efficient in a distributed-memory setting. Each block is first reduced locally on a process.
When the blocks have been reduced to the best of a \majorRevision{process's} capabilities,
the unreduced columns are sent to the next process to the left. These communication and computation steps are performed 
until all columns are completely reduced. The persistence pairs of the diagram can then 
be extracted from the columns and rows of the reduced matrix.
% For the same dataset, the DIPHA output will always be identical, regardless of the number of
% processes used for the computation.
\majorRevision{\emph{DIPHA}} offers good parallel
speedups, using only multi-process (MPI) parallelism (no multi-threading), and allows for 
the analysis of larger datasets than
\julien{anterior work.}
% previously possible.

Another
\julien{distributed}
method was introduced by Ceccaroni et al. in
\cite{ceccaroniDistributedApproachPersistent2024a}.
However, their parallelism %among compute nodes
is limited to the concurrent processing of multiple\julien{, smaller} datasets
\julien{(i.e., at least one full dataset per process)}.
%%It aims at computing persistent homology in a distributed-memory setting. However, their method deals 
%%with batches of datasets. Each dataset individually is processed on a single computing unit. Parallelism 
%%in that approach refers to concurrently processing multiple datasets at once. This is not our 
%%setting:
On the contrary, we aim in this paper at processing one voluminous
dataset on multiple compute nodes. % a distributed-memory computer.
An approach recently introduced by Nigmetov et al. \cite{cadmus} also uses
distributed-memory parallelization to produce a persistence diagram. The algorithm 
combines spatial and range partitioning by computing first a local reduction of the 
data and then switching to a global reduction. However, this algorithm relies on 
persistent co-homology while our algorithm uses homology. As stated 
by Nigmetov et al., experiments suggest that persistent homology may be more efficient
for computing the persistence diagram. Experiments
that we performed
% using
\julien{with}
\majorRevision{\emph{DIPHA}}
% to compute diagrams
using both homology and co-homology confirm this. At the time of
writing this paper, no public implementation has been found for the approach of 
Nigmetov et al.
To our knowledge, \majorRevision{\emph{DIPHA}} is thus the only public implementation for 
distributed-memory computation of persistent homology. 

\majorRevision{In contrast to \majorRevision{\emph{DIPHA}} and the approach of 
Nigmetov et al. \julienRevision{(which both consider a distribution of the data
by scalar value)}, our algorithm employs a spatial distribution of the input
data and entirely obviates the need to compute the \julienRevision{global}
boundary matrix. This feature
constitutes a practical advantage, especially in contexts where data distribution has already been 
determined upstream in the analysis pipeline. \julienRevision{Also, the}
% The
reliance on discrete Morse Theory
yields challenges that intrinsically differ from the solutions offered by the two other 
approaches.}

\section{Overview}
\label{sec_overview}

This section provides an overview of our approach.
First, the global order of
\julien{vertices}
% each vertex
is computed. This step is called
\emph{Array Preconditioning}. In a distributed-memory setting, a global order is necessary for 
comparing vertices owned by different processes. 
The computation is done in parallel on multiple processes in three steps:
we start by creating % Pierre : tu est déjà dans un "first" %%first, we create
locally a vector for all vertices of the elements
to sort (comprised of the scalar value of the vertex, its global identifier and the rank of the process 
that owns it). \majorRevision{A distributed sort is then performed on the vector using psort} \cite{psort}. This particular 
implementation was chosen because it is lightweight, modifiable and relatively 
efficient. \majorRevision{\emph{DIPHA}} also uses
this implementation to construct its boundary matrix. Each process can then compute the global order 
of the elements that are present locally following the distributed sort. 
Finally, the global orders are sent back to the owner of the corresponding vertices. 
For simplices of higher dimension, comparisons are performed using the lexicographic 
comparison on their global vertex orders (\autoref{sec_inputData}) .

Second, the discrete gradient of the input data is computed by using the algorithm described
by Robins et al. \cite{robins_pami11}, which is embarrassingly %trivially
parallel for both shared- and distributed-memory contexts \cite{leguillou_tech24}.

Third, the critical simplices are extracted from the gradient and sorted in the step 
\emph{Extract \& sort}.

Fourth, the diagrams $\diagram_0$ and $\diagram_{2}$ are computed by processing the 
unstable and stable sets of the 1-saddles and $2$-saddles of $f$
and applying a self-correcting pairing algorithm to extract the pairs of the diagram \autoref{sec_extr-sad}.

Next, the diagram $\diagram_{1}$ is computed from the 
unpaired $1$- and $2$-saddles using our novel algorithm
"DistributedPairCriticalSimplices" \autoref{sec_sad-sad}.

Finally, the classes of infinite persistence are extracted by collecting the remaining, 
unpaired critical simplices.

\majorRevision{As 
\julienRevision{the input to}
the steps \emph{Extract \& sort}, $\diagram_0$, $\diagram_{2}$ and 
$\diagram_{1}$ are 
% applied exclusively to 
\julienRevision{the}
critical simplices, our overall algorithm is output-sensitive 
\julienRevision{(the number of critical simplices directly determines the 
output size, the number of persistence pairs)}. 
For grids of 
% the 
identical dimension, the number of simplices involved in these 
steps depends on the input scalar fields, potentially ranging from very few to 
a large number. The more pairs are present in
the output persistence diagram, the greater the workload.
Consequently, two datasets of the same dimension may yield persistence 
diagrams of markedly different sizes, with correspondingly varying computation times.}

\section{Extremum-Saddle Persistence Pairs}
\label{sec_extr-sad}
\begin{figure*}
  \centering
  \includegraphics[width=\linewidth]{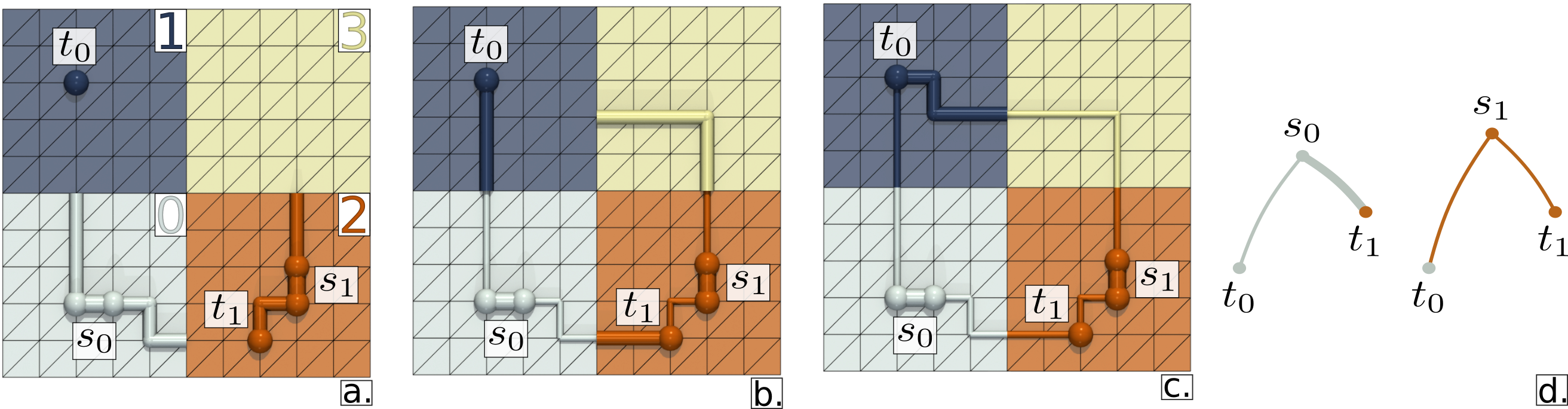}
  \caption{Overview of our algorithm for extremum-saddle pairing for $\diagram_0$.
  First, the unstable set is computed from the vertices of the critical $1$-simplices $s_0$
  and $s_1$ (\julien{sub-}figure a, the  two thick edges with their vertices)
by following the gradient. When the computations of sets reach ghost simplices,
a message is sent to the relevant process
  to notify it to resume computation on its domain (\julien{sub-}figure b, the
newly computed part of the set is thicker). Rounds of computations and
communications
  are performed until all sets are computed (\julien{sub-}figure c).
  The set is then collapsed into the distributed graph $\graph{0}$ following the rules 
  stated in \autoref{sec_distrGraph} (\julien{sub-}figure d). The ownership of
extremum $t_0$ is given to
  process 0 as it is the process with the lowest rank identifier that has a set that 
  ends in $t_0$.
  The ownership of extremum $t_1$ is given to process 2 as it has an unstable set started 
  in $\domain_2$, at $s_1$, that ends in $t_1$.
  The graph is then processed with a distributed Union-Find structure to produce the persistence 
  pair in $\diagram_0$. The thicker arc ($s_0$, $t_1$) corresponds to the
computed
  \julien{persistence}
  pair.}
  \label{fig_extr-sad}
\end{figure*}

In this section, we will describe the different modifications to
\julien{PairExtremaSaddles}
% PairExtremumSaddle
(see \autoref{alg_pairExtremumSaddle}) that we
\julien{contributed}
% implemented
to obtain a distributed-memory version.
This algorithm is applied to compute both $\diagram_0$ and $\diagram_{2}$. 
An overview of the algorithm is provided in \autoref{fig_extr-sad}.

\subsection{Stable and unstable sets computation}
\label{sec_distributedUnstableSets}

The unstable sets of all critical \majorRevision{$1$-simplices} are computed as follows:
a v-path is extracted from each vertex of each critical $1$-simplex.
Then, two possibilities arise: a critical vertex is encountered (i.e. a local minimum), ending the
computation there for this unstable set, or a ghost vertex is encountered, in which 
case a message is stored to be later sent to the process owning the ghost 
vertex so that the computation can resume there later. Once all computations either are  
completed or have generated a message, all processes exchange their 
stored messages and resume their computations on their block. These successive computation and 
communication steps run until no messages are sent on any process during a communication 
round.
For stable sets, the computation is similar but is applied on $2$-simplices as 
\majorRevision{the} start of the set and $3$-simplices as \majorRevision{the final simplex of the set}. The gradient is followed in 
reverse to emulate the dual gradient without explicitly computing it.

\subsection{Distributed extremum graph construction}
\label{sec_distrGraph}

The previous step computes the stable and unstable sets. Now we 
% need
% to 
collapse those sets into the distributed extremum graphs $\graph{i}$ with $i
\in \{0, 2\}$.
\julien{In the remainder, we focus on the case $i=0$, the case $i=2$ being
symmetric.}
The nodes of the graph are the saddles and extrema of the
\julien{previously computed sets (\autoref{sec_distributedUnstableSets}).}
% stable and
% unstable sets.
The arcs of the graph represent the v-paths connecting \julien{these} critical
simplices.
A triplet of the graph refers to the three nodes $(\sigma, t_0, t_1)$,  where $\sigma$
is a critical saddle and $t_0$ and $t_1$ extrema linked to $\sigma$ by v-paths.

Nodes of this graph may be located on different processes.
% It is also possible that
% several sets end at the same extremum.
We therefore establish a few additional
rules to fit the definitions of \julien{$\graph{0}$} to a distributed-memory
setting. \majorRevision{These rules were established to enable the computation of the next 
  step of the algorithm (i.e. the self-correcting pairing).}
\begin{itemize}
  \item Saddles are present on only one process. \majorRevision{This arises from the manner in 
which representatives are computed in \pierreRevision{distributed memory:}
%% Pierre : pour alléger %% a distributed-memory setting:
the process is 
initiated at a saddle and subsequently proceeds from one extremum to another, 
thereby eliminating the need for ghost saddles. It is necessary to determine which 
process holds the next extremum of the representative computation, whereas this 
requirement does not apply to saddles.}
\item A saddle node in
\julien{$\graph{0}$} is owned
by a process $p$ if its associated critical simplex is \julien{exclusively}
owned by $p$ (\autoref{sec_distributedModel}). \majorRevision{This ownership 
  limits %% Pierre : the
  computation and communication at this step, as graph saddles are 
created where their originating saddles in the triangulation already exist.}
\item Extrema, however, can be present 
on multiple processes but they are owned by only one and are ghost on other processes.
\majorRevision{This design choice was driven by the need to determine to which process 
to resume the representative computation in the \pierreRevision{self-correcting} algorithm. With the presence 
of ghosts, a necessary link is made between processes.}
\item The ownership of an extremum node is determined as follows. If an extremum
simplex $e$ is \julien{exclusively} owned
by the process $p$ (\autoref{sec_distributedModel}) and if there exists a
saddle simplex $s$ also \julien{exclusively} owned
by $p$ such that one of its unstable sets terminates in $e$, then the extremum
node
\julien{of $\graph{0}$}
associated to $e$ is \emph{owned} by $p$. Otherwise, the node associated to $e$ is 
\emph{owned} by the process with the lowest rank identifier that owns a saddle
node
whose unstable set ends in $e$.
\majorRevision{The ownership of the extrema was designed to limit the number of communications.
However, as the construction of the graph and the \pierreRevision{self-correcting} algorithm are saddle-centric, 
it is more \pierreRevision{convenient} %% Pierre (subjectif) %% practical
to ensure that an 
% extrema 
\julienRevision{extremum}
is on the same process of at least one of the saddles 
it is linked to in the graph, hence the ownership design.}
\end{itemize}

\autoref{fig_extr-sad} shows 
an example of a distributed extremum graph and applied ownership rules. The
local graph on a process $p$ is noted \julien{$\localGraph{0}{p}$}. The ghosted
local graph
of a process $p$ is noted \julien{$\ghostedGraph{0}{p}$}. An  extremum node is
called \emph{at the interface}
of two processes $p$ and $q$ if it is owned by either $p$ or $q$ and is a ghost on the 
other process.

The computation of the collapse of the sets to build a graph is fairly
straightforward: once all
the sets are computed, the processes possess lists of all sets ending on a local extremum 
they own with regard to their domain. Each process will then determine which process 
is the owner of the extremum in the graph and send back to the owner of the 
originating saddle the extremum and its new ownership. It will also send to the new owner of the 
node a list of all the processes on which the extremum is a ghost with regard to the graph.
Each process will then receive and build the parts of
\julien{its local graph.}
% the graph present locally.

\subsection{Self-correcting distributed pairing}
\label{sec_selfcorrecting}

\begin{figure*}
  \centering
  \includegraphics[width=\textwidth]{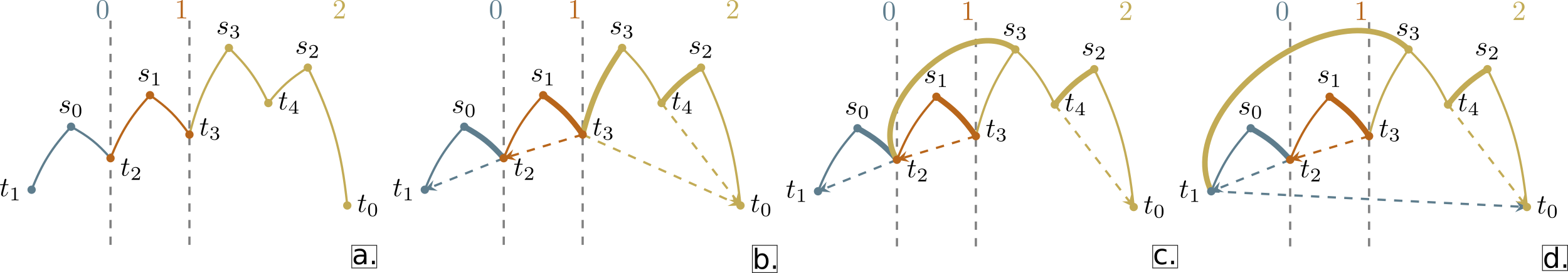}
  \caption{Illustration of our self-correcting pairing algorithm on an example of $\graph{0}$. 
  \julien{Sub-f}igure a represents a distributed
  \julien{extremum}
  graph on 3 processes, 0 (in 
blue), 1 (in orange)
  and 2 (in yellow). Extrema at the interface of two processes intersect with the 
  dashed grey lines separating the domain of the processes. \julien{Sub-f}igure 
b represents 
  the first computation iteration. Each process computes on its local graph without 
  taking other processes into consideration. On process 2, $s_2$ is paired with $t_4$ 
  (shown here by a thicker arc) and $t_0$ is made the new representative of $t_4$
  (shown here by a dashed arrow). Then the pairing of $s_3$ is performed: $s_3$
  can be paired with either $t_3$ or the representative of $t_4$, $t_0$. Since $t_3$
  is 
  \julien{higher}
%   bigger 
  than $t_0$, $s_3$ is paired with $t_3$. On process 1, $s_1$ is also 
paired 
  with $t_3$. After the computation, processes exchange data regarding 
relevant, \julien{shared} 
pairings. 
  Process 2 will tell process 1 it created 
  the pairing $(s_3, t_3)$ with $t_0$ as the new representative, because $t_3$ 
  is owned by process 1. Process 1 will tell process 2 about its owned pairing 
  ($(s_1, t_3)$, with $t_2$ as the new representative) because it knows that $t_3$ is present 
  as a ghost on the graph of process 2. Process 2, upon receiving this message, will 
  assess that the message is correct and its pairing $(s_3, t_3)$ is wrong.
  It will compute a new pairing:  $(s_3, t_2)$ (as $t_2$ is the new representative
   of $t_3$) and 
  tell process 1 about the pairing (\julien{sub-f}igure c). Process 1 will 
update $t_2$ in the pairing to its 
  representative,  $t_1$. As $t_1$ is owned by process 0, process 1 will tell it about the pairing.
  Process 0, knowing that $t_1$ is unpaired, will accept the pairing $(s_3, t_1)$\majorRevision{. Then, it will} 
  update the representative of $t_1$ to $t_0$\majorRevision{. Finally, it will }send back the information to 
  process 2 that the actual correct pairing is $(s_3, t_1)$ as shown in \julien{sub-f}igure d.}
  \label{fig_selfcorrecting}
\end{figure*}

\pierre{We now have to build the $\diagram_{0}$ pairs from the local}
  \eve{graph $\ghostedGraph{0}{p}$} \pierre{(and similarly
  for $\diagram_{2}$).
This step was originally performed sequentially in DMS since its execution time was
negligible compared to others.}
However, in a distributed-memory setting,
\majorRevision{a sequential execution would most likely greatly hurt the 
overall performance} as it would prevent
any speedup \pierre{on multiple nodes} \majorRevision{for this particular step.
A detailed analysis of the execution time associated with DMS reveals that, for both 
$\diagram_0$ and $\diagram_2$, this 
\julienRevision{sequential}
stage can account for up to $8.6\%$ of the total 
runtime for Magnetic Reconnection, one of our test dataset\julienRevision{s} 
(see \autoref{sec_datasets}). 
According to Amdahl's law, the theoretical upper bound on
\pierreRevision{the multi-node speedup} 
%% relative to multi-node  acceleration
under these conditions is $\frac{1}{0.086} \approx 11.6$. 
This limitation would therefore impose a significant constraint on 
the attainable efficiency of the algorithm as a whole.
% Furthermore,
Also, and most importantly,
sequential execution on one node would require to gather the intermediate data 
(i.e\julienRevision{.,} the graph computed in \autoref{sec_distrGraph}) on the 
node. This would \pierreRevision{introduce additional communications
  and may} create an additional 
memory bottleneck in an algorithm that \pierreRevision{has already
  high memory requirements}.
%is already highly memory-intensive. %% Pierre : éviter confusion avec memory-bound
}

We thus have to design a distributed pairing algorithm. 
\majorRevision{The new algorithm heavily relies on algorithm \autoref{alg_pairExtremumSaddle},
  but introduces several components to face the challenges brought by
  \pierreRevision{distributed-memory parallelism.}
  %% an execution in a distributed-memory setting. 
We incorporated a self-correction mechanism, enriched the representation of extrema 
with additional data, removed the path compression mechanism, employed maps and local 
identifiers in place of global identifiers, and introduced
inter-process \pierreRevision{communications} %exchanges 
where necessary. The following paragraphs provide a detailed account of these 
modifications.}

We use the idea of comparing saddles to see which is the oldest
  (or the youngest), but restrict the algorithm to the distributed
  extremum graph created at the previous step. 
Our algorithm design is also
inspired by the self-correcting mechanism of the
\pierre{parallel (multi-threaded)} %parallelized 
procedure \emph{PairCriticalSimplices}, \pierre{computing
  $\diagram_{1}$ in  the original DMS algorithm.
This introduces} % Although
  incorrect pairings \pierre{resulting in % may introduce
  extra computations. Such overhead is however handled 
in parallel, ultimately \majorRevision{resulting in}
overall performance gains (see \autoref{sec_results}).} %in a distributed-memory setting.
Comparisons to the original DMS algorithm in \autoref{fig_compDMS} \majorRevision{and \autoref{table_gain}} \pierre{also} show that our 
method incurs limited overhead. 
\pierre{Moreover,} we expected other strategies relying on a more direct approach
\pierre{(i.e. without incorrect pairings)} to induce 
too much synchronizations and idle time to perform well.

A key difference from the original DMS algorithm is that the representative of an 
extremum will now store two pieces of information: the representative and the identifier 
of the saddle that assigned the representative to the extremum. This enables, when 
computing the pairing of a saddle $\sigma$, to stop the computation 
of the representatives of its extrema nodes $t_0$ and $t_1$ when the loop reaches representatives assigned 
by saddles older than $\sigma$ (as this would not occur in sequential). 
The path compression mechanism of DMS mentioned in \autoref{sec_originalDMS} (see \autoref{alg_pairExtremumSaddle}) is no longer 
applied as the resulting representatives may be false,
\pierre{leading} to potentially 
incorrect computations for other pairings during the computation of other representatives. 
When a wrong pairing is detected 
\julien{by saddle comparison,}
% through comparison of saddles, 
the computation of the 
representatives of its original triplet is \julien{re-}started 
% again 
from the 
beginning using \autoref{alg_distProcessTriplet}.

\begin{algorithm}
  \caption{DistributedProcessTriplet}\label{alg_distProcessTriplet}
  \begin{algorithmic}[1]
      \small
      \REQUIRE A triplet $(\sigma, t_{\julien{0}}, t_{\julien{1}})$ of 
$\localGraph{0}{p}$ on process $p$
      \ENSURE A temporary pair of $\diagram_{0}$
      \FOR[Compute the representatives]{$i$ in $\{0,1\}$}
      \STATE $r_i \gets findRepresentative(t_i, \sigma)$
      \STATE $r_iPaired \gets Pair(r_i) \ne \emptyset$
      \STATE $r_iInvalid \gets r_iPaired$ and $\sigma < Pair(r_i)$ 
      \IF{$r_iInvalid$}
        \STATE $r_iPaired \gets false$
      \ENDIF
      \ENDFOR    
      \IF{$r_0 \notin \ghostedGraph{0}{p}$ or $r_1 \notin \ghostedGraph{0}{p}$} 
        \STATE Send $(\sigma, r_0, r_1)$ to owner
      \ENDIF
      \IF{\eve{$r_0 == r_1$}}
        \IF[$\sigma$ should not be paired]{\eve{$Pair(\sigma) \neq \emptyset$}}
        \STATE \eve{$r \gets Pair(\sigma)$}
        \STATE \eve{$Representative[r] \gets r$}
        \COMMENT{re-initialize the representative}
        \STATE \eve{$Pair(\sigma), Pair(r) \gets \emptyset, \emptyset$}
        \COMMENT{remove the invalid pair}
        \ENDIF
      \ELSE
        \IF{($r_0 < r_1$ or $r_0Paired$) and $!(r_1Paired)$}
          \STATE swap $r_0$ and $r_1$ data
        \ENDIF
        \IF{$!(r_0Paired)$}
          \IF[remove the invalid pair]{$r_0Invalid$}
            \STATE $\sigma_k \gets Pair(r_0)$
            \STATE $ Pair(\sigma_k) \gets \emptyset$
          \ENDIF
          \STATE $Pair(\sigma), Pair(r_0)  \gets r_0, \sigma$ \COMMENT{Add the new pair}
          \STATE \eve{$Representative[r_0] \gets (r_1, \sigma)$}
          \IF{$r_0$ is at the interface of $p$ and process $q$}
            \STATE Send $(\sigma, r_0, r_1)$ to process $q$
          \ENDIF   
          \IF[Recompute the invalid $\sigma_k$]{$r_0Invalid$}
            \IF{$\sigma_k \in \localGraph{0}{p}$}
              \STATE DistributedProcessTriplet($\sigma_k$)
            \ELSE 
              \STATE Send recomputation signal to the owner of $\sigma_k$
            \ENDIF
          \ENDIF
        \ENDIF        
      \ENDIF
      \end{algorithmic}
  \end{algorithm}

Here is a description of the overall self-correcting
\pierre{distributed} algorithm. 
A practical example is shown in \autoref{fig_selfcorrecting}.
First, each process executes \emph{DistributedProcessTriplet} (See \autoref{alg_distProcessTriplet})
for all its triplets (in a sorted manner,
as it proved to be more efficient even though it is not required). 
The messages to be sent are stored until all computations are performed\majorRevision{,} 
then all processes exchange their messages. 
For each message $(\sigma, m_0, m_1)$ received, the process will first detect if it is a 
recomputation (encoded by $(\sigma, -1, -1))$ and will trigger it. If it is not a recomputation, 
the process will update the representatives of the message to 
$(\sigma, Representative(m_0, \sigma), Representative(m_1, \sigma))$.
If the newly computed representatives belong to another process, then the message 
is passed along to that other process. Otherwise, the process detects if the 
message should trigger a correction \pierre{through} saddle comparison. If so, the pairs and the representatives are 
updated and a recomputation is triggered for the invalid saddle. 
This cycle of communications and computations is repeated until no messages are sent on 
any process during a communication round.

The original DMS algorithm used vectors to \pierre{store} %implement
several variables relative to 
critical simplices \eve{in its} \emph{PairCriticalSimplices} algorithm. These vectors were 
defined for all simplices of the triangulation, even though they were used only 
for critical simplices.
This allowed for fast \pierre{memory accesses}, % of memory,
as the index of the simplex in the triangulation
was equal \pierre{its} %to the
index in the vectors. In a distributed-memory setting, this is no longer possible:
the global number of simplices will most
likely prevent the \pierre{memory allocation} %creation
of such large vectors. We therefore reduced the size of the vectors to the number of local 
critical simplices and used maps to 
% obtain the 
convert a global simplex index to its 
index in \julien{these} vectors.

\subsection{Shared-memory parallelism}
\label{sec_sharedMem}

The computations of stable and unstable sets are straightforwardly
processed with multiple threads \pierre{in each MPI process} %trivially parallelized with threads
as all set computations are independent. % the computation of each set is independent from another.
The communications are however performed only by the \eve{OpenMP primary} thread. The construction of 
the distributed graph is similarly parallelized.

The self-correcting pairing algorithm is not multi-threaded
\pierreRevision{within each MPI process}. 
\majorRevision{Although feasible in principle, we made the choice not to explore shared-memory 
parallelism for this particular step. Our expectations were that
\pierreRevision{multi-thread} %%multithreaded 
parallelization would yield only limited performance gains, an assessment 
informed by the performance results reported for a comparable algorithm by 
Smirnov et al. \cite{smirnov17}. Furthermore, as shown by 
Guillou et al.  \cite{guillou_tech22} (Table 3, Appendix C), the computation of 
$\diagram_0$ and $\diagram_{2}$ with DMS is often limited in terms of computation time 
with regard to the other procedures. 
Since this step \pierreRevision{already benefits from
  distributed-memory parallelism,} %is already parallelized within a distributed-memory setting, 
we considered that the additional complexity introduced by thread-level parallelization 
would most likely outweigh the potential performance benefits.}

$\diagram_0$ and $\diagram_{2}$ being completely independent, we assigned each 
diagram computation to an OpenMP task that can itself generate threads in a nested 
manner.

\section{Saddle-Saddle Persistence Pairs}
\label{sec_sad-sad}
In this section, we will describe the different modifications to the original \emph{PairCriticalSimplex} 
(\autoref{alg_pairCriticalSimplex}) and \emph{PairCriticalSimplices} 
(\autoref{alg_pairCriticalSimplices})
algorithms necessary for the efficient distributed computation of $\diagram_{1}$.
%% Pierre : qd on implémente un algorithme, on obtient du code
%% that we implemented to obtain a distributed-memory algorithm. 

\subsection{Distributed-memory parallel algorithm}
\label{sec_basic}
\begin{figure*}
    \includegraphics[width=\linewidth]{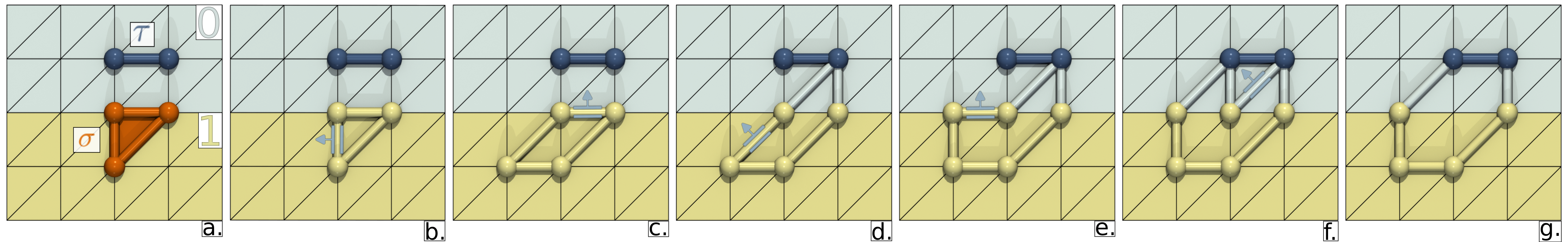}
    \caption{\julien{Distributed} homologous propagation. 
    \julien{Sub-f}igure a shows two 
critical simplices: $\sigma$, the death saddle, and 
    $\tau$, the birth saddle. $\sigma$ is located on process 1 (light yellow) and $\tau$ is located on 
    process 0 (light blue). Initially, the boundary of $\sigma$ is only located 
on process 1 (\julien{Sub-f}igure b).
    New edges are added to the boundary that are owned by 
    process 1 (\julien{Sub-f}igure c). Then new edges are added to the boundary 
that are ghost for process 1 (\julien{Sub-f}igure d) and a message is sent to 
process 0
    so that it adds the edges to its local boundary of $\sigma$. On process 1, a new piece of information is stored regarding the boundary:
    the highest edge for process 0 (equal to the highest edge of the two ghost edges). As the global highest edge is still
    located on process 1 (gray arrow), the computation continues on process 1 
(\julien{Sub-f}igure d) and additional ghost edges are added 
(\julien{Sub-f}igure e).
    After adding two new 
    ghost edges (\julien{Sub-f}igure f), the highest edge is now located on 
process 0. Process 1 will send the computation token
    to process 0. Upon receiving it, process 0 will resume the computation and 
propagate the boundary (\julien{Sub-f}igure f).
    Finally, the propagation ends in $\tau$ and the pair ($\sigma$,$\tau$) is created on process 0.}
    \label{fig_propagation}
  \end{figure*}

Our algorithm relies on a new data structure: the global-local
boundary. For a $2$-simplex $\sigma$, 
this structure is composed of two elements: the set of edges 
% associated to the 
\julien{of the boundary initiated in $\sigma$ (called in the remainder, for 
simplicity, the \emph{boundary} of $\sigma$)},
% \emph{boundary} \julienof $\sigma$, 
which are owned by the current process $p$ 
(called 
\emph{local boundary}), and the highest boundary edges of all processes
containing a part of the boundary of  $\sigma$ 
%for each process it is present on
(called the \emph{global boundary}).
%% Pierre : dans la version précédente, j'avais l'impression que
%% c'était une structure distribuée avec une seul arête dans chaque
%% processus 
The local boundary is identical to the boundary of the original 
DMS algorithm. The global boundary is updated by other processes all through the computation. 
At any given time, on a given process $p$, the highest edge $\sigma$ of its local boundary is always lower or
equal to the highest edges reported by the global boundaries of the other processes. 

A central idea in our distributed-memory algorithm is the notion of
\emph{computation token}. Each $2$-simplex 
$\sigma$ for which a 
\julien{homologous}
propagation needs to be computed is associated
with a token. At any time, the token of each propagation is present in only one
process. %, There is only one for all processes.
Only the process owning the token is allowed to propagate the boundary of $\sigma$.
This means that\julien{, taken individually,} each propagation is 
\julien{carried out}
% processed 
%performed
sequentially.
% \julien{(i.e., only one active process at a given time)}. 
However, all the propagations are 
computed in parallel similarly to the %what was done in a
shared-memory context. 

Here is a description of our algorithm \emph{DistributedPairCriticalSimplex} as defined in 
\autoref{alg_distPairCriticalSimplex} and illustrated %shown
in \autoref{fig_propagation}.
This algorithm revisits  \emph{PairCriticalSimplex} 
\autoref{alg_pairCriticalSimplex}  in order to compute a distributed homologous
propagation for an unpaired critical 2-simplex $\sigma$.  
For an unpaired critical $2$-simplex $\sigma \in \domain_p$, with $p$ a process,
its local boundary is propagated by following the same rules as %the original
\emph{PairCriticalSimplex}. 
However, when a ghost edge, owned by process $q$, needs to be added to the 
global-local boundary of $\sigma$, a message will be sent to %is generated for %to be later sent to 
$q$ so that $q$ %to tell it to
adds this particular edge to its local boundary of $\sigma$.
\julien{The propagation can trigger a merge between two boundaries. Local 
boundaries}
% If the propagation triggers a merge between
% two boundaries, the local boundaries 
% \julien{to the considered process}
are merged similarly to %the
%original
\emph{PairCriticalSimplex} %algorithm 
(\autoref{alg_pairCriticalSimplex}, l.15-17).
% The two 
\julien{Global}
% global
boundaries are merged by keeping the highest of the two
edges for each process. A message will then 
be sent to notify all the relevant processes that a merge has occurred and should be performed by them as well.
As soon as the highest edge of the local boundary is no longer the highest edge in the global boundary, the 
propagation on $p$ is stopped. Then, the computation token will
be sent to the process 
owning the highest edge in the global boundary, to resume the propagation on its block.

At the end of the computation, the pair $(\tau, \sigma)$ is stored on the process 
that owns $\tau$, as the boundary of another propagation may reach $\tau$ and a comparison 
between the two originating $2$-simplices may be required. The process that owns 
$\sigma$ does not have to be aware of which simplex completed the pair. %It is unnecessary

\emph{DistributedPairCriticalSimplex} hence generates two types of messages: 
computation tokens and boundary updates. Boundary updates
correspond to either a merge order between two global-local boundaries, an addition of an edge to a local boundary or an update of the highest edge 
in a global boundary. The received boundary updates 
have to be performed in a particular order, to ensure two properties: (i) updates 
from potentially multiple processes with regard to one particular propagation need 
to be received and processed in the same order they were created in, (ii) updates from 
one process with regard to potentially multiple propagations need also 
to be received and processed in the same order they were created in.
Other orders may result in an incorrect outcome.
For (i), the property can be ensured by following a round-by-round 
design with alternating communication and computation steps. Processing the 
boundary updates sequentially for each process will ensure property (ii). 
However, messages sent by different processes can be processed in any order as long as 
they involve different propagations.

\begin{algorithm}
  \caption{DistributedPairCriticalSimplex}\label{alg_distPairCriticalSimplex}
  \begin{algorithmic}[1]
      \small
      \REQUIRE An unpaired critical $2$-simplex $\sigma$
      \ENSURE A temporary pair of $\diagram_{1}$
      \IF{$\globalLocalBoundary{\sigma} == 0$}
        \STATE \emph{addEdge}($\globalLocalBoundary{\sigma}, \partial \sigma$)
      \ENDIF
      \WHILE{$\globalLocalBoundary{\sigma} \ne 0$}
        \STATE $\tau \gets$ max$(\localBoundary{\sigma})$
        \STATE $\tau_{p} \gets$ max$(\globalBoundary{\sigma})$
        \IF[$\tau_p$ is the highest edge]{max$(\tau, \tau_{p}) == \tau_{p}$}
        \STATE \emph{UpdateMaxGlobal}$(\sigma, \tau, \globalBoundary{\sigma})$
        %        \STATE Send computation token of $\sigma$ to $p$
        %% Pierre : sauf erreur de ma part, tu ne fais pas l'envoi dès maintenant 
        \STATE Mark computation token of $\sigma$ for sending to $p$
        \STATE \textbf{return}
        \ENDIF \COMMENT{$\tau$ is the highest edge}
        \IF{$\tau$ is not a critical simplex } 
          \STATE \emph{addEdge}($\globalLocalBoundary{\sigma}, \partial(Pair(\tau))$
        \ELSE
          \IF{$Pair(\tau) == \emptyset$}
            \STATE addPair$(\sigma, \tau)$
            \hfill\COMMENT{$\tau$ is unpaired}  
            \STATE \emph{UpdateMaxGlobal}$(\sigma, \tau, \globalBoundary{\sigma})$
            \STATE \textbf{break}
          \ELSE 
            \STATE $\sigma_{\tau} \gets Pair(\tau)$
            \hfill\COMMENT{$\tau$ has already been paired to $\sigma_{\tau}$}
            \IF{$\sigma_{\tau} < \sigma$}
              \STATE \emph{MergeGlobalLocalBoundaries}$(\sigma, \sigma_{\tau})$
            \ELSE
            \STATE addPair$(\sigma, \tau)$
            \hfill \COMMENT{$\sigma$ is older and the true death of $\tau$}
            \STATE \emph{UpdateMaxGlobal}$(\sigma, \tau, \globalBoundary{\sigma})$            
            \STATE $Pair(\sigma_{\tau}) \gets \emptyset $
            \STATE PairCriticalSimplex$(\sigma_{\tau})$
            \hfill \COMMENT{Resume for $\sigma_{\tau}$}
            \ENDIF
          \ENDIF
        \ENDIF
      \ENDWHILE
      \end{algorithmic}
  \end{algorithm}

The overall algorithm \emph{DistributedPairCriticalSimplices} 
% defined in 
(\autoref{alg_distPairCriticalSimplices})
revisits  \emph{PairCriticalSimplices} (\autoref{alg_pairCriticalSimplices})
% using multiple 
\julien{with}
distributed processes, and following this round-by-round
design.  
%is structured as follows: all
All the propagations are first computed locally using multi-thread parallelism.
Once all propagations are either completed or their computation token needs to be sent to 
another process, the communications start. First, the boundary updates are exchanged and processed.
Then the computation tokens are exchanged and the new propagations are computed in parallel using threads.
These rounds of communications and computations are performed until all
critical $2$-simplices are paired.
%TODO: following algorithm necessary to be shown?
\begin{algorithm}
  \caption{DistributedPairCriticalSimplices}\label{alg_distPairCriticalSimplices}
  \begin{algorithmic}[1]
      \small
      \REQUIRE Set $C_{2}$ of unpaired critical $2$-simplices \majorRevision{$\sigma$}
      \REQUIRE Set $C_{1}$ of unpaired critical $1$-simplices \majorRevision{$\tau$}
%%      \REQUIRE Integer $n$: number of processes   
      \ENSURE \majorRevision{The persistence diagram $\diagram_{1}$, a collection 
      of pairs $(\sigma, \tau)$}
%%      \STATE \# parallel for
      \FOR{$j \in C_{2}$ {\it in parallel (multi-threading)}}
        \STATE DistributedPairCriticalSimplex$(\sigma_j)$
      \ENDFOR
      \WHILE{\eve{Global number of terminated propagations} $\lt |C_2|$}   
        \STATE \COMMENT{Perform global boundary updates}
        \STATE Send boundary updates to other processes
        \STATE Receive boundary updates from other processes
        \STATE Update \pierre{received boundaries}
        \STATE \COMMENT{Resume computations with tokens}
        \STATE Send computation tokens to other processes
        \STATE Receive computation tokens from other processes
%%          \STATE \# parallel for
        \FOR{all received tokens $\sigma$ {\it in parallel (multi-threading)}}
          \STATE DistributedPairCriticalSimplex$(\sigma)$
        \ENDFOR        
      \ENDWHILE
      \FOR[Extract pairs from boundary computation]{$j \in C_{1}$}
        \STATE $\diagram_{1} \gets \diagram_{1} \cup (\sigma_j, Pair(\sigma_j))$  
      \ENDFOR       
      \end{algorithmic}
  \end{algorithm}

Similarly to \autoref{sec_selfcorrecting}, our implementation reduced the size of data structures 
and vectors to the local number of critical simplices and used maps to
convert a global simplex index to its index in vectors.

\subsection{Anticipation of propagation computation}
\label{sec_anticipation}
\begin{figure*}
    \includegraphics[width=\linewidth]{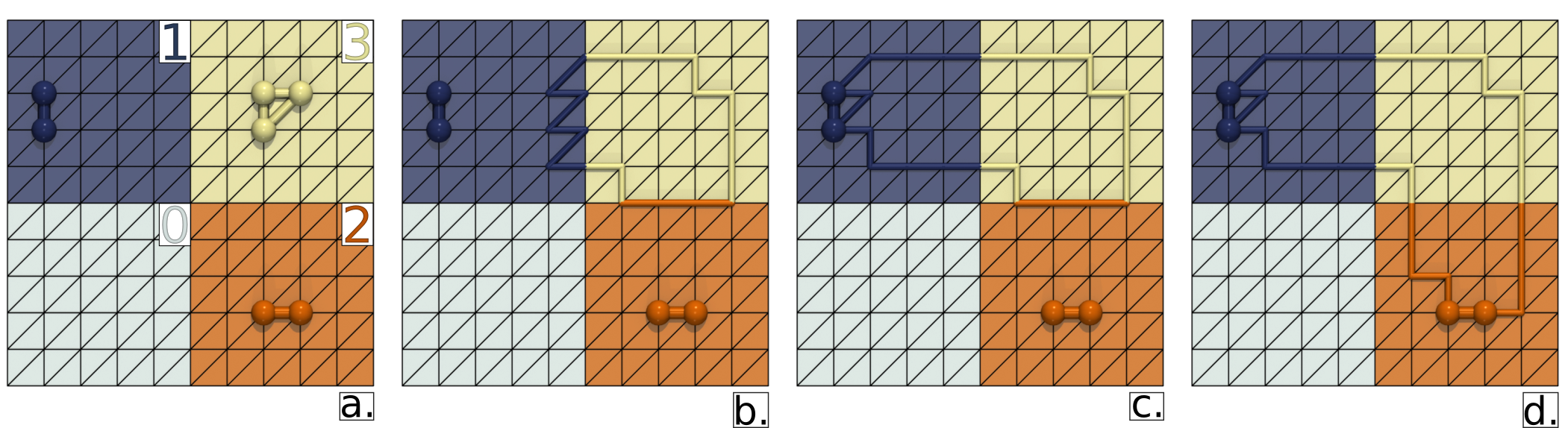}
    \caption{Example of anticipation of propagation computation for 4 processes (process 0 in light blue, process 1 in 
    dark blue, process 2 in orange and process 3 in light yellow). A critical triangle
    (in yellow) is the starting point of the distributed 
    \julien{homologous}
    propagation. The boundary is 
    propagated as described in \autoref{alg_distPairCriticalSimplex}, however the 
    computation token is not sent when the global
    highest edge of the boundary becomes located on process 1 
(\julien{Sub-f}igure b). Instead, the propagation is 
    continued on process 3 until the number of propagation iterations
    %steps
    (i.e. while loop iterations in \autoref{alg_distPairCriticalSimplex})
    %steps
    reaches a predefined 
    counter. Then, the computation token is sent to the process owning the 
    global highest edge (here, process 1). The boundary is propagated on process 
1 (\julien{Sub-f}igure c)
    until an unpaired critical edge is reached \julien{(thick, dark, blue 
edge)}. At this point, the global highest edge is located 
    on process 2 (orange). The 
    \julien{blue critical}
    edge is therefore not paired with the critical triangle
    and the computation token is sent to process 2 that resumes the propagation on its 
    domain. When reaching its own critical edge, as the global highest edge is still located 
    on the domain of process 2, the propagation ends here and the pair is created on process 2
    (\julien{Sub-f}igure d). Instead of having to send the computation token 
back and forth between
    process 1, 2 and 3 to produce the final pairing, the anticipation of propagation enables 
    to exchange the token only twice.
    }
    \label{fig_anticipation}
  \end{figure*}

The algorithm presented in the previous section presents a critical flaw: in the worst 
case scenario, for boundaries stretched out on multiple processes, the 
maximum
% \julien{token}
may 
change 
% \julien{switch}
process 
every time a simplex is added to the propagation, resulting in repeated exchanges of the 
computation token.
This may generate an extremely high number of communications. Anticipating this back 
and forth is possible by changing slightly the original algorithm as shown in
\autoref{fig_anticipation}. Instead of stopping 
the computation and sending the computation token to another process as soon as %when
the highest edge of the boundary is located on another process, we further the computation regardless 
on the current process, until either the number of propagation
iterations (i.e., while loop iterations in 
\autoref{alg_distPairCriticalSimplex}) %steps
reaches 
a predefined counter (arbitrarily equal to $0.01\%$ of the number of triangles of $\domain_p'$) or until an unpaired critical simplex $c$ is reached.
Only then is the computation token sent to the process that owns the highest edge.
Not pairing the potential simplex $c$ ensures that the propagation
never expands %goes
too far, leading to potentially incorrect pairs, which would have been
difficult to detect and correct afterwards. % required additional testing, synchronizations and messages to identify and correct.

%%%%% Pierre : pas d'élément nouveau a priori, et paragraphe précédent
%%%%% trop court pour nécessiter un résumé 
%% This change greatly reduces the number of computation and communication rounds 
%% by computing much more at once than the basic algorithm, while ensuring no simplex 
%% is wrongly paired with a propagation that goes too far.

\subsection{Overlap of communication and computation}
\label{sec_commthread}

There are two limitations to the previous algorithm that we want to address in this 
section: thread idle time and cost of communication. 
Indeed, at the end of each computation round, when waiting for all work to 
complete, there is often just a few propagations being computed, resulting in 
significant idle time when using many threads. We aim at reducing this idle time 
by triggering a communication round before all computations are finished.
On the other hand, the cost of communication can be reduced by
%% Pierre : on ne le résoud pas complètement a priori % addressed by
effectively overlapping communications with computations at the MPI level. 

A dedicated \emph{communication thread} can solve both these problems. We preserve the round-by-round 
structure for the communications to ensure that the update of 
global-local boundaries is processed in the right order (see \autoref{sec_basic}),
but %is performed by a dedicated communication thread in each MPI process. 
the communication rounds are now triggered by the communication threads. 
In each MPI process, the communication thread sends and receives messages, updates boundary data 
and creates one OpenMP task for each propagation, while
the other %% Pierre OpenMP
{\em compute threads} process the propagations. 

Even if we aim at starting earlier each communication round,
  these %The communication rounds
are not triggered %Pierre : pour éviter répétition % started
as soon as one message can be sent. Making the communication thread wait 
\majorRevision{until reaching a threshold}
and sending multiple messages in one MPI communication at once
limits indeed the number of communications. It ensures that the MPI layer is
not overloaded with numerous messages, and limits the 
number of OpenMP 
% Pierre : je précise OpenMP car il y a aussi des opérations atomiques
% en MPI désormais
atomic operations performed by the communication thread: these atomic
operations are required for a correct synchronization with the compute threads.
Messages will only be sent \majorRevision{if the number of messages waiting to be sent by the current %on a
process is above the following threshold. 
It is set dynamically to increase reactivity as the computation progresses.}
At first, it is equal to $0.01\%$ of the local number of unpaired $2$-simplices.
Then, at every round of communications, it is updated using the remaining 
global number of unpaired critical $2$-simplices to add reactivity to the communications.
\majorRevision{This value of $0.01\%$ was determined empirically. In the case where
there are no tasks left to be computed within the current process, messages are sent 
regardless of whether the threshold was reached.}
The communication thread also performs the update of global-local boundaries. This can 
be done in parallel of propagation computation as updating global-local boundaries from 
the current round will not interfere with the computation of propagations from 
previous rounds. 
% This 
\julienRevision{It}
is because the updates are not directly related to the current 
computation tokens.

The compute threads can now 
continuously process the local and incoming propagations through a task pool 
filled by the communication thread, harnessing more efficiently the intra-node multi-core parallelism. 
The idle time of the threads is therefore significantly reduced as compute threads 
no longer have to wait at the end of each computation round and 
the cost of communication is effectively 
hidden with the overlap of communication with computation as usual with
a communication thread \cite{denisMPIOverlapBenchmark2016a, hager2011}.

\section{Results}
\label{sec_results}
For the following results, we rely on Sorbonne Université's supercomputer, %MCMeSU-std. 
MCMeSU, which contains 48 nodes of 32 cores each. Each
node is composed of 2 AMD EPYC 7313 Milan CPUs with 256GB of
\julien{RAM.}
% memory.
The nodes are interconnected with Mellanox Infiniband. In our tests, 
\julien{we use up to 16 nodes (512 cores total), with one}
% we rely on one 
MPI process 
% with 
\julien{and}
32 threads per node to minimize MPI communications and synchronizations as well
as the memory footprint. %% Pierre : non définie jusqu'à présent : For the \emph{Overlap} version,
When using a communication thread (see
  \autoref{sec_commthread}), we rely on 31 compute threads only.
%32 threads are used in total (1 communication thread and 31 computation 
% threads).
\julien{Our algorithm is implemented in C++ with \pierre{MPI+}OpenMP within TTK 
\cite{ttk17, ttk19}. The correctness of our implementation was 
% systematically 
% tested
checked
for all test datasets, by comparing our outputs against those generated 
by DMS (which were already compared to \majorRevision{\emph{DIPHA}}'s for 
triangulated
voxel data,
% 3D domains,
see \cite{guillou_tech22} for more details).}
Tests of strong and weak scaling
% benchmarks 
are conducted to study the performance of our algorithm.
The preconditioning time
\julien{for TTK's distributed triangulation (\autoref{sec_distributedModel})}
% of the TTK triangulation
% will not be studied nor 
\julien{is not}
accounted 
for in this paper \julien{(it is negligible for regular grids
\cite{leguillou_tech24}).
Specifically, when reporting execution times, we consider that the data is
already distributed among the nodes, in the form of ghosted blocks
(\autoref{sec_distributedModel}), which is a standard input for analysis
pipelines in distributed environments.
Only} the
execution time of our algorithm and its direct preconditioning
% will be 
\julien{are}
measured. Our algorithm is also compared to the original DMS algorithm 
as well as to \majorRevision{\emph{DIPHA}}.

\subsection{Datasets}
\label{sec_datasets}

The performance of our software has been evaluated using multiple datasets, 
selected to demonstrate a broad spectrum of cases. These datasets are sourced from 
publicly available repositories \cite{openSciVisDataSets, ttkData}. 
\majorRevision{Datasets from the OpenSciVis %dataset
  collection can be prepared using the script available in the reproducible example (See \autoref{sec_contributions}).}

\begin{itemize}
    \item Backpack: a CT scan of a backpack 
    \julien{containing various items.}
%     with items. 
    This \julien{acquired} dataset is spatially
    imbalanced with regard to its number and location of topological features, making it 
    a good test case for workload balancing.
    \item Isabel: a simulation of hurricane Isabel that hit the east coast of the USA in 
    2003. This dataset is very smooth and will result in a persistence diagram with 
    few but significant pairs.
    \item Elevation: a synthetic dataset of the altitude within a cube, with a unique maximum
    at one corner of the cube and a unique minimum at the opposite corner. This is 
    a pathological case with only 
    \julien{one class of infinite persistence in $\diagram_0$.}
%     the infinite pair in its persistence diagram.
    \item Wavelet: a synthetic dataset of 3D wavelets on a cube. This dataset is quite
    smooth and symmetric, which should result in a very good workload balance.
    \item Isotropic pressure: a direct numerical simulation of forced isotropic 
    turbulence. This dataset is originally very big ($4096^3$) and was down-sampled.
    \item Magnetic reconnection: A single time step from a computational simulation of magnetic 
    reconnection, showing interactions between magnetic fields. This dataset is extremely 
    noisy.
    \item Synthetic truss: A simulated CT scan of a truss with defects. This dataset 
    possesses very rich and symmetric topological features.
    \item Random: a synthetic dataset of a random field on a cube. This dataset 
    showcases a pathological case \julien{with a high number of spatially 
evenly distributed persistence pairs.}
%     where the number of persistence pairs of its diagram
%     is high and the pairs are spatially evenly distributed.
\end{itemize}

For the strong scaling benchmarks, all datasets were resampled to $512^3$ via trilinear 
interpolation, except for Random, that was resampled to $512^2 \times 256$
as the execution time for this dataset is particularly long for $512^3$ with all tested softwares, 
making it unpractical to manage. This smaller size still makes Random the dataset with the 
longest execution time, as it is our worst case scenario.

For the weak scaling benchmarks, the size of the input 
\julien{(number of vertices)}
doubles each times the number 
of nodes %cores
doubles
\julien{(by doubling the number of vertices along one, alternating, dimension)}.
\eve{The initial size on one node is the same as
  %the size of
  the strong  scaling one %benchmark
  \julien{($512^2 \times 256$ for Random, $512^3$ for the others)}.
%   ($512^3$ and $512^2 \times 256$ for Random).
The datasets were re-sampled in different ways depending on the size of the 
original data: Isabel, Backpack, Magnetic Reconnection have been 
up-sampled, whereas Synthetic truss and Isotropic Pressure have been down-sampled. 
Random was generated for its biggest weak scaling case and then down-sampled 
to smaller datasets. Elevation was generated for each size, so that it always 
has only one pair in its diagram. Due to its symmetry, Wavelet was generated for the largest weak 
scaling case and then resized by being cut in two along each
dimension \julien{alternatively}.}
%% Pierre : pas trop envie de parler de l'impact sur le workload ici
%% , as its symmetry ensured a proportional division of the workload.}

\subsection{Performance improvements}

\begin{figure}
    \centering
    \includegraphics[width=\linewidth]{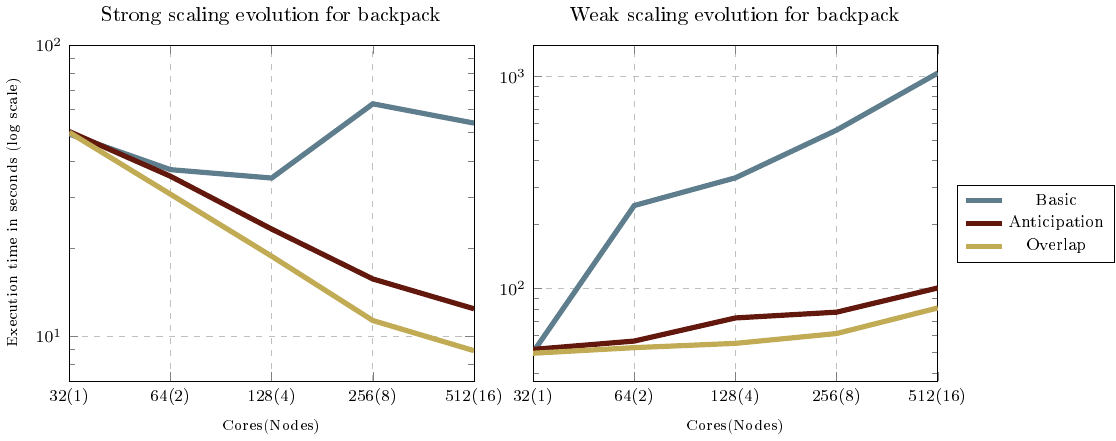}
    \caption{Performance impact of the
      different $\diagram_{1}$ versions on the overall DDMS execution time
      for Backpack. %improvements of our algorithm.
      \emph{Basic} corresponds to the 
    first version for  $\diagram_{1}$  described in
    \autoref{sec_basic}, \emph{Anticipation} %corresponds
    to the 
    second one %algorithm as
    described in \autoref{sec_anticipation}, and 
    \emph{Overlap} to the final one %corresponds to the
    %last iteration
    described in \autoref{sec_commthread}.}
    \label{fig_improvements}
\end{figure}

We start by assessing the performance improvements of our different
versions for computing $\diagram_{1}$, %%algorithms on the backpack dataset:
%%\autoref{fig_improvements} shows performance results of our 
%%$\diagram_{1}$ %%algorithms on the backpack dataset:
namely: \emph{Basic}, the initial version (see \autoref{sec_basic}); 
%basic algorithm of our algorithm,
\emph{Anticipation}, that implements 
the anticipation of computation for $\diagram_{1}$ (see \autoref{sec_anticipation})
and \emph{Overlap} that iterates on \emph{Anticipation} and adds %to the computation of $\diagram_{1}$ 
the overlap of communication and computations thanks to the
  communication thread (see \autoref{sec_commthread}).
As shown in \autoref{fig_improvements}, 
the anticipation of computation dramatically improves the overall DDMS
performance, % of the algorithm,
making \emph{Anticipation} 6 times faster than \emph{Basic} on 16
nodes %process
in strong scaling and over 12 
times in weak scaling. \emph{Overlap} also improves the performance, by adding reactivity to the 
execution. On 16 nodes, it reduces the overall execution time by 20\% in weak 
scaling and 28\% in strong scaling. These results validate and
  justify our modifications, which are hence necessary to efficiently deploy such TDA
algorithms on multiple nodes. 

\begin{figure}
    \centering
    \includegraphics[width=\linewidth]{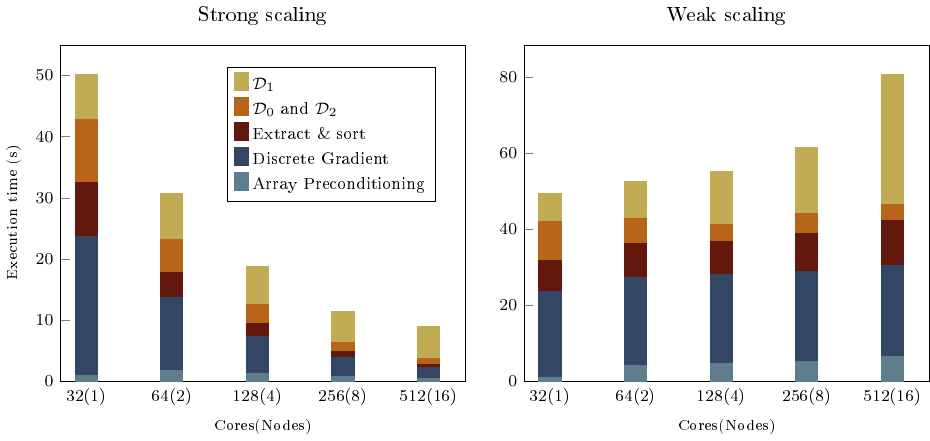}
    \caption{Execution time of each step of DDMS, for strong (left) 
      and weak (right) scalings for Backpack.}
      %% Pierre : déjà mis dans le texte 
      %% The "Array Preconditioning" phase corresponds to the 
      %% computation of the global ordering of vertices and "Extract \& Sort" to the
      %% extraction and local sort of critical simplices for all dimensions
      %% (see \autoref{sec_overview}).
    \label{fig_barChart}
\end{figure}

From now on, we will only consider the \emph{Overlap} version,
whose detailed execution profile is presented in 
\autoref{fig_barChart}. % shows the execution time of this version.
%% "Array Preconditioning" corresponds to the computation of global order of vertices and
%% "Extract \& sort" to the extraction of critical simplices of all dimension and their 
%% local sort.
For both strong and weak scalings, the \emph{Array Preconditioning}
step, which  corresponds to the computation of global order of
vertices (see \autoref{sec_overview}), is quite short %small
and minority. \majorRevision{In strong scaling, \emph{Array Preconditioning} is 
marginally
more expensive on two nodes than on one, after which execution 
time decreases and ultimately 
% surpasses 
\julienRevision{outperforms}
% the 
single-node performance.
% This 
\julienRevision{It}
is due to the cost of the 
% additional 
\julienRevision{extra}
computation steps described in 
\autoref{sec_overview},
mainly the preparation of the vector and the computation of the global order from the sorted 
vector. Despite this increase, the computation of this step remains a minority.}

% Pierre : je dirai pas vraiment "négligeable" pour 16
% processus en weak scaling %negligible.
The discrete gradient step scales very %really
well as expected, which is 
% the main 
\julien{an important}
source of overall performance gains. 
The computations of $\diagram_{0}$ and $\diagram_{2}$ scale also well,
in both \julien{weak} and strong scaling.
\majorRevision{The following step, \emph{Extract \& Sort},} corresponds to the extraction and local sort of critical
simplices for all dimensions (see \autoref{sec_overview}).
\pierreRevision{It is} performed independently \julien{on} each
process %.
\majorRevision{\pierreRevision{and} %It
  also scales well.}
Only the computation \pierreRevision{time} of $\diagram_{1}$
\julien{(which is the most intensive in terms of time complexity)}
\pierreRevision{remains constant in strong scaling and}
scales 
\julien{unfavorably}
% poorly
in %%both strong and
weak scaling.
But, thanks to our successive improvements presented in
\autoref{sec_sad-sad}, there is no strong performance loss.
\majorRevision{%%In strong scaling, the performance of D1 remains constant.
Taking advantage of the embarrassingly parallel steps of our algorithm,} 
we still manage to obtain overall significant
performance gains when increasing the number of nodes. %leads to %produces a good overall acceleration.
%% % does not scale very well
%% and remains %stays
%% relatively constant as the number of processes increases. 
%% In the weak scaling case, the computation of $\diagram_{1}$ eventually becomes 
%% quite significant in the overall execution for 16 processes.
%%
%% However, thanks to the good scalability of the other steps, we 
%% always obtain performance improvements when increasing the number of
%% nodes. %leads to %produces a good overall acceleration.
%%
%% In conclusion, the overall scaling performance heavily relies on the discrete 
%% gradient scalability as expected. Other steps also show good scalability, except for 
%% the computation of $\diagram_{1}$, that has much more limited scalability.

\begin{figure}
    \centering
    \includegraphics[width=\linewidth]{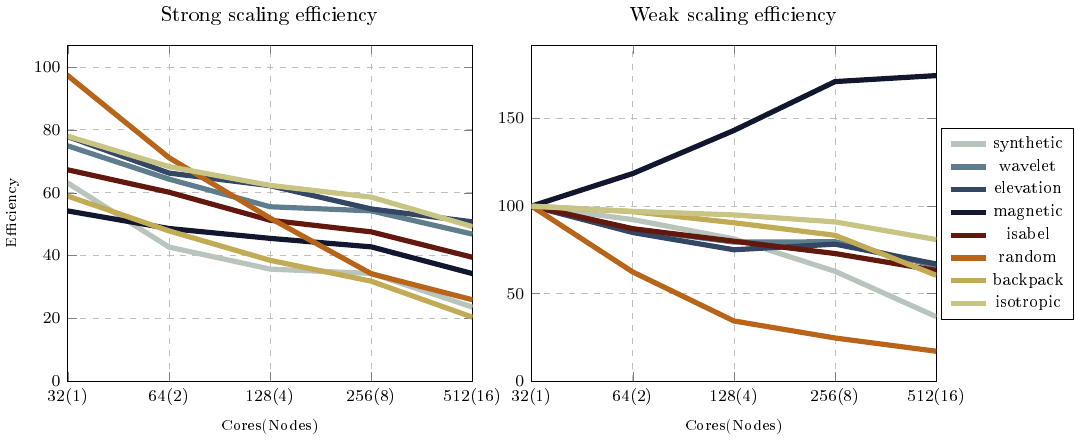}
    \caption{Parallel efficiency of DDMS for strong (left) and weak (right)
scaling.}
%     benchmarks.}
    \label{fig_strongWeakScaling}
\end{figure}

\subsection{Strong scaling}
\label{sec_strongScaling}
The results for all datasets in strong scaling are
\julien{shown} in \autoref{fig_strongWeakScaling} (left) in
terms of \julien{parallel} efficiency with respect to the execution on one core
\julien{(i.e., speedup divided by the number of cores, in percentage)}. The
execution times are also available in \autoref{fig_compDMS}.
With the exception of Random, the efficiencies of all datasets fit within the 
range of 55\% to 80\% on one node and of 20\% to 50\% on 512 cores.
This shows the scalability of our approach.
% is quite good results.
Though the efficiencies decrease as the number 
of cores increases, \autoref{fig_compDMS} shows that the execution times continue 
to decrease even on 512 cores, with most datasets eventually requiring less 
than 20 seconds
% to execute
on 512 cores. \majorRevision{These performance results demonstrate that the 
benefits of multi-node execution extend beyond increased memory availability: 
they also encompass the acceleration of computations that could already be performed 
on a single node.}

Random behaves a bit worse than the other datasets, presenting
%it sees
the biggest drop in efficiency: % and goes
from %having
the best efficiency on one node (close
to 100\%) to one of the worst one on 512 cores (close to 26\%).
This is explained by the output-sensitivity of our algorithm 
\majorRevision{(see \autoref{sec_overview})}. 
%more work there will be to compute  during the execution.
Another factor is the spatial placement of the birth and death 
of a pair within the dataset. The further apart they are, the longer the computation 
will be. Random is one of the noisiest of our datasets (with Magnetic Reconnection 
and Synthetic Truss), however, unlike those two datasets
its pairs are %it is
evenly distributed.
Consequently, the birth and death
% are
\julien{tend to be}
further apart spatially, requiring more work
and more communications. % to be computed.
This leads to very good efficiency on one node,
but this quickly becomes a performance issue % problem
as the number of nodes increases, leading to such
an efficiency drop.

\subsection{Weak scaling}
\label{sec_weakScalin}

We define the weak scaling efficiency for $p$ processes %is defined
as $e_p = \frac{t_1}{t_p} \times 100$, with $t_1$ and $t_p$ being the execution times 
on 1 and $p$ nodes.
The weak scaling results for all datasets are shown in \autoref{fig_strongWeakScaling} 
(right). \eve{The weak scaling efficiency is better than the strong scaling one for most datasets.
This is partly due to the fact that doubling the dataset size through 
re-sampling often results in less than a twofold increase
\julien{in the number of critical simplices, and hence}
in computational workload \julien{(given the output sensitivity of our
algorithm)}.}
For most datasets, the efficiency is %very good and
in the range of 35\% to 80\% on 16 nodes, which again shows the
scalability of our approach. There are however two exceptions: Random and Magnetic Reconnection.
For Random, the efficiency eventually drops lower to 17 \% for %. This
%is due to
the same reasons as in strong scaling: its pairs are numerous and spatially 
stretched out. For Magnetic Reconnection, the efficiency largely exceeds 100\%.
This is due to the up-sampling of the original dataset that barely multiplies the 
number of pairs by a factor of 1.6 between 1 and 16 nodes. This is most likely 
because the topological features are already numerous and unevenly distributed 
across the dataset. While %But
this also applies to
%However, it is not enough to justify such results, as
other datasets, 
%face the same problem,
such as Isabel, 
% Another factor is at play here: the number  of pairs produced.
the other specificity of Magnetic Reconnection is that it produces the most pairs out of all the 
datasets on $512^3$ (35 millions). The execution time
% of computing
\julien{for}
$\diagram_{0}$, $\diagram_{1}$
and $\diagram_{2}$ is therefore substantial compared to the computation of other steps, such as 
the gradient. As the number of nodes %cores
increases, even though the size of the dataset 
increases, each process actually computes less and less pairs
leading to lower execution times.
%% causing an additional decrease in %the
%%execution time.
The same applies for Isabel, but has no impact on
% the same phenomenon occurs, but it is not visible in 
the overall execution time, as the computation of $\diagram_{0}$, $\diagram_{1}$
and $\diagram_{2}$ is originally negligible compared to the gradient computation.

\subsection{Performance comparison}
\label{sec_perfComparison}

\begin{figure}
    \centering
    \includegraphics[width=\linewidth]{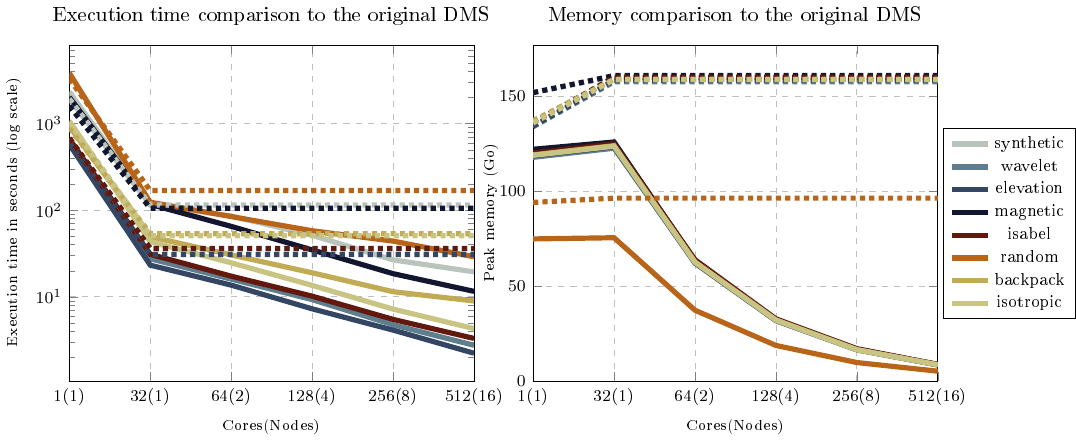}
    \caption{Comparison between DMS (dotted lines) and DDMS (full lines)
    in terms of execution time (left) and
    \julien{per-node}
    peak memory footprint (right).}
    \label{fig_compDMS}
\end{figure}

\begin{table}
\begin{center}
\resizebox{\linewidth}{!}{%
\begin{tabular}{|r||rrrrrrrr|r|}
\hline
\#cores & Synthetic & Wavelet & Elevation & Magnetic & Isabel  & Random  & 
Backpack & Isotropic  & Average\\
\hline     
1 &   -20.38  & \textbf{2.14}   &  \textbf{2.95} &   -26.11 &  \textbf{2.15}    & -16.54    &    -3.98  &  -3.95  & -8.00  \\
\hline
32 &  -0.83  & \textbf{23.39}  &  \textbf{24.87}  &   -10.86 & \textbf{14.68}   & \textbf{26.87}    &  \textbf{7.55}  & \textbf{14.71} & \textbf{12.5}  \\
\hline
\end{tabular}
}
\end{center}

\caption{\majorRevision{\textnormal{Gain of DDMS over DMS (in percentage).
A positive gain \julienRevision{(bold)} indicates that DDMS outperforms DMS in terms of execution speed.}}}
\label{table_gain}
\end{table}

\textit{Comparison with DMS:} In \autoref{fig_compDMS} are shown the execution time %(left)
and
\julien{per-node}
peak memory footprint %(right)
of our DDMS algorithm %(full lines)
compared to the original DMS algorithm on one core and then one node. %(dotted lines).
\majorRevision{In \autoref{table_gain} are shown the gains of DDMS over DMS for executions 
on 1 and 32 cores. A positive gain indicates that DDMS outperforms DMS in terms of execution speed.
On 1 core, the gain is on average $-8\%$. 
The datasets characterized by larger workloads \julienRevision{(in terms of
number of critical simplices)}, namely Random,
Synthetic Truss, and
Magnetic Reconnection, exhibit the lowest gains. This suggests that \pierreRevision{gain} 
decreases as workload size increases. 
Then, on 32 cores, the gain increases for 
all datasets, yielding comparable} execution times of both algorithms, the
distributed DDMS algorithm executing slightly faster than its shared-memory counterpart
for all datasets, except Synthetic Truss and Magnetic Reconnection.
The DDMS extra cost for these two datasets %Synthetic Truss and
%Magnetic Reconnection
is due to changes 
in the algorithm (for example, the removal of arc collapse in the $\diagram_{0}$ and
$\diagram_{2}$ computations in \autoref{sec_selfcorrecting}) or to more costly data structures 
(\autoref{sec_distrGraph}),
% that enable the
\julien{enabling}
MPI execution. % in a distributed memory setting.
The overhead is however very 
limited \majorRevision{on average} and executing DDMS on two nodes already outperforms DMS for all datasets.

In terms of peak memory footprint, DDMS uses significantly less memory for all datasets. This 
is due to our reduction of vector size to the number of critical simplices as mentioned in 
\autoref{sec_selfcorrecting} and \autoref{sec_basic}. This allowed DDMS to produce an 
overall smaller footprint \julien{on one node}.

\begin{figure}
    \centering
    \includegraphics[width=\linewidth]{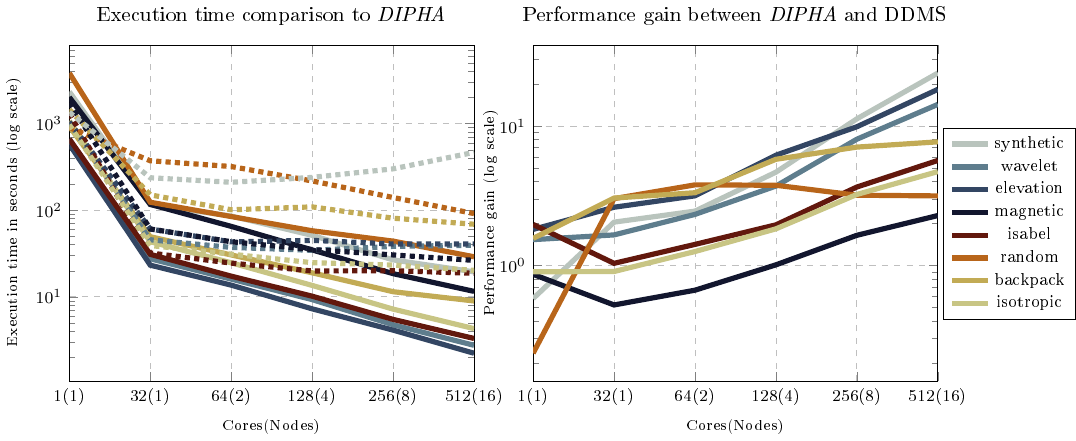}
    \caption{Comparison based on execution time (left) %comparison
      between \majorRevision{\emph{DIPHA}} (dotted lines) and 
    DDMS (full lines) and performance gain (right) for a strong scaling 
    setting. The performance gain on 
    a given %certain
    number of cores is defined as $t_{\majorRevision{\emph{DIPHA}}}/t_{DDMS}$, 
    %$\frac{t_{Dipha}}{t_{DDMS}}$,
    with $t_{\majorRevision{\emph{DIPHA}}}$ and $t_{DDMS}$ the execution times of
\majorRevision{\emph{DIPHA}} and
    DDMS respectively.
    A performance gain higher than 1 means that DDMS is faster than 
\majorRevision{\emph{DIPHA}}.
   }
    \label{fig_compDipha}
\end{figure}

\textit{Comparison with \it{DIPHA}:} We now compare our algorithm to
\majorRevision{\emph{DIPHA}},
\julien{to our knowledge}
the only
publicly available MPI implementation (without multithreading) for persistence diagram
computation. %a distributed-memory context.
%%\majorRevision{\emph{DIPHA}} is executed using only MPI processes (no multithreading).
The \majorRevision{\emph{DIPHA}} execution time is measured using the built-in \emph{benchmark} 
mode and corresponds 
to its total execution without the I/O time. 
%% The comparison of our performance to \majorRevision{\emph{DIPHA}} is shown in 
%% \autoref{fig_compDipha} and \autoref{fig_compDiphaMem}. In
%% \autoref{fig_compDipha}, the left figure shows the execution time for
%% a strong scaling  
%% benchmark with DDMS and \majorRevision{\emph{DIPHA}}. On the right is 
%% shown the acceleration factor between DDMS and \majorRevision{\emph{DIPHA}}. 
%%
We start by comparing the execution times in strong scaling on
\autoref{fig_compDipha}. 
On one core,
%% Pierre : se déduit de la phrase suivante : 
%% \majorRevision{\emph{DIPHA}} shows a better performance on several datasets (Isotropic pressure, Random, Magnetic Reconnection and Synthetic Truss).
DDMS outperforms \majorRevision{\emph{DIPHA}} %show a better performance
only on the smoother datasets (Elevation, Wavelet, Isabel and Backpack), as 
these can really harness the preconditioning of the discrete gradient to 
speed up the rest of the computation.
%TODO: rappeler raisons? (pierre)
On multiple % cores and
nodes, DDMS scales much %overall
better %than \majorRevision{\emph{DIPHA}}
and hence outperforms \majorRevision{\emph{DIPHA}} for all datasets starting
from 4 nodes. 
%% eventually on 16 nodes,
%% the performance of DDMS is better for all datasets.
Notice than on one and two nodes, only one dataset out of eight (Magnetic Reconnection) is more
efficiently processed by \majorRevision{\emph{DIPHA}}. 
Moreover, considering both execution times and scaling, the \julien{worst} case
dataset
for  \majorRevision{\emph{DIPHA}}
(Synthetic Truss) scales relatively well for DDMS, 
%% Three datasets in particular draw 
%% our attention: Synthetic Truss, Random and Magnetic Reconnection.
%% Synthetic Truss is a dataset that does not scale well with \majorRevision{\emph{DIPHA}}: 
%% the execution time quickly plateaus and even increases. For DDMS, this dataset shows 
%% average performance, scaling up relatively well. DDMS therefore performs well for cases 
%% that are problematic for \majorRevision{\emph{DIPHA}}.
whereas the \julien{worst} case for DDMS (Random) is always processed
faster (up to
\julien{$\times 3$)}
% 3x)
by DDMS on more than one core.
%% our algorithm still quickly performs better 
%% than \majorRevision{\emph{DIPHA}}, eventually executing 3 times faster.
%%
%% One dataset shows \majorRevision{\emph{DIPHA}} to perform better than DDMS: Magnetic Reconnection.
%% For 1 to 64 cores, \majorRevision{\emph{DIPHA}} outperforms our algorithm but DDMS eventually
%% catches up and surpasses its performance.
Finally, the average
% performance gain
\julien{speedup for}
% over
all
datasets is
% approximately
\julien{around $\times 8$}
% 8x
on 512 cores, showing
\julien{a substantial}
% an overall large
%significant
performance
gain of DDMS
% \julien{on}
over %our algorithm with regard to
\majorRevision{\emph{DIPHA}}.

\begin{figure}
    \centering
    \includegraphics[width=\linewidth]{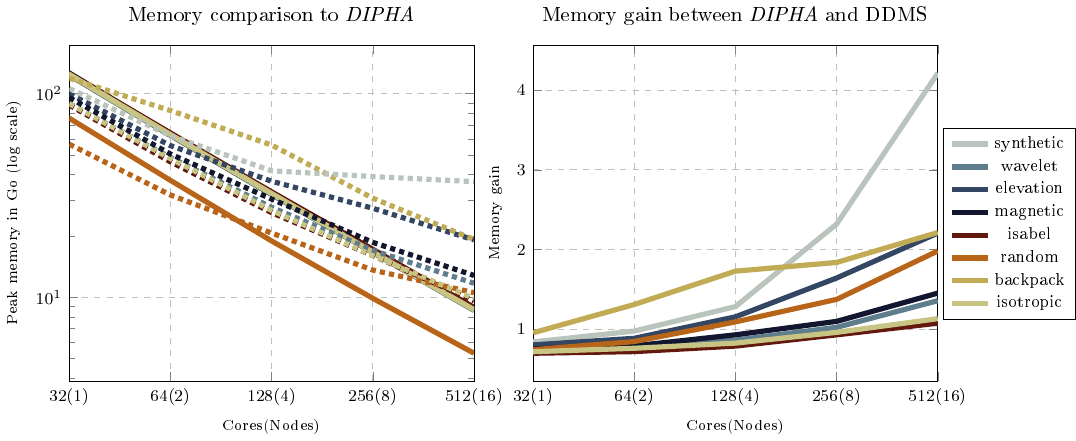}
    \caption{Maximum over all nodes of the
    \julien{per-node}
    peak memory footprint
%     per node
    (left)
for \majorRevision{\emph{DIPHA}} (dotted lines) 
    and DDMS (full lines) and memory gain (right) for a strong
    scaling setting. The memory gain on 
    a given %certain
    number of nodes is defined as $m_{\majorRevision{\emph{DIPHA}}}/m_{DDMS}$, %    
% $\frac{m_{Dipha}}{m_{DDMS}}$,
    with $m_{\majorRevision{\emph{DIPHA}}}$ and $m_{DDMS}$ the maximum over all node of the
    \julien{per-node} peak
    memory footprint for  \majorRevision{\emph{DIPHA}} and
DDMS respectively. A memory gain higher than 1 means that \majorRevision{\emph{DIPHA}} uses more
memory than
DDMS.}
    \label{fig_compDiphaMem}
\end{figure}

\autoref{fig_compDiphaMem} compares %shows
the memory consumption of both
\julien{approaches}.
% codes.
%% The left figure shows the maximum over all node of peak memory footprint for DDMS (full lines) 
%% and \majorRevision{\emph{DIPHA}} (dotted lines). On the right is shown the memory factor between DDMS and 
%% \majorRevision{\emph{DIPHA}}, defined similarly as the speedup factor ($\frac{m_{Dipha}}{m_{DDMS}}$, with 
%% $m_{Dipha}$ and $m_{DDMS}$ the maximum over all node of peak memory memory footprint on a node respectively of \majorRevision{\emph{DIPHA}} and 
%% DDMS). A factor higher than 1 means that \majorRevision{\emph{DIPHA}} uses more memory than DDMS. 
On the fewest number of nodes, \majorRevision{\emph{DIPHA}} requires a lower
\julien{footprint}
%   memory space
%Pierre : voir explication à la fin du paragraphe uses less memory
than DDMS for all datasets but 
the memory scalability is much better for DDMS. 
Hence, on the largest number of nodes the peak memory footprint of DDMS ends
up being smaller than
\julien{that of}
\majorRevision{\emph{DIPHA}}.
DDMS divides indeed by almost two its memory usage (with a small overhead due to the ghost cells) 
every time the number of cores is multiplied by two
(\autoref{sec_distributedModel}). %This is due to the way the data is
% distributed.

\eve{For \majorRevision{\emph{DIPHA}}, its data distribution is more conducive to memory imbalance as 
the final reduced columns are stored on the process that completed their 
reduction, meaning that some process may store a significantly higher number 
of reduced columns than other processes.
This imbalance is more likely as the number of processes increases,
leading to \majorRevision{\emph{DIPHA}} having a larger per-node peak memory footprint than DDMS.}
% Regarding
% data
%% Pierre : comme c'est déséquilibré pour
%% Dipha, peut être que sur l'ensemble des noeuds Dipha consomme en
%% tout moins de mémoire, mais on considère bien ici le max du peak
%% sur tous les noeuds

\subsection{Example}
\label{sec_bigExample}

\begin{figure}
    \centering
    \includegraphics[width=\linewidth]{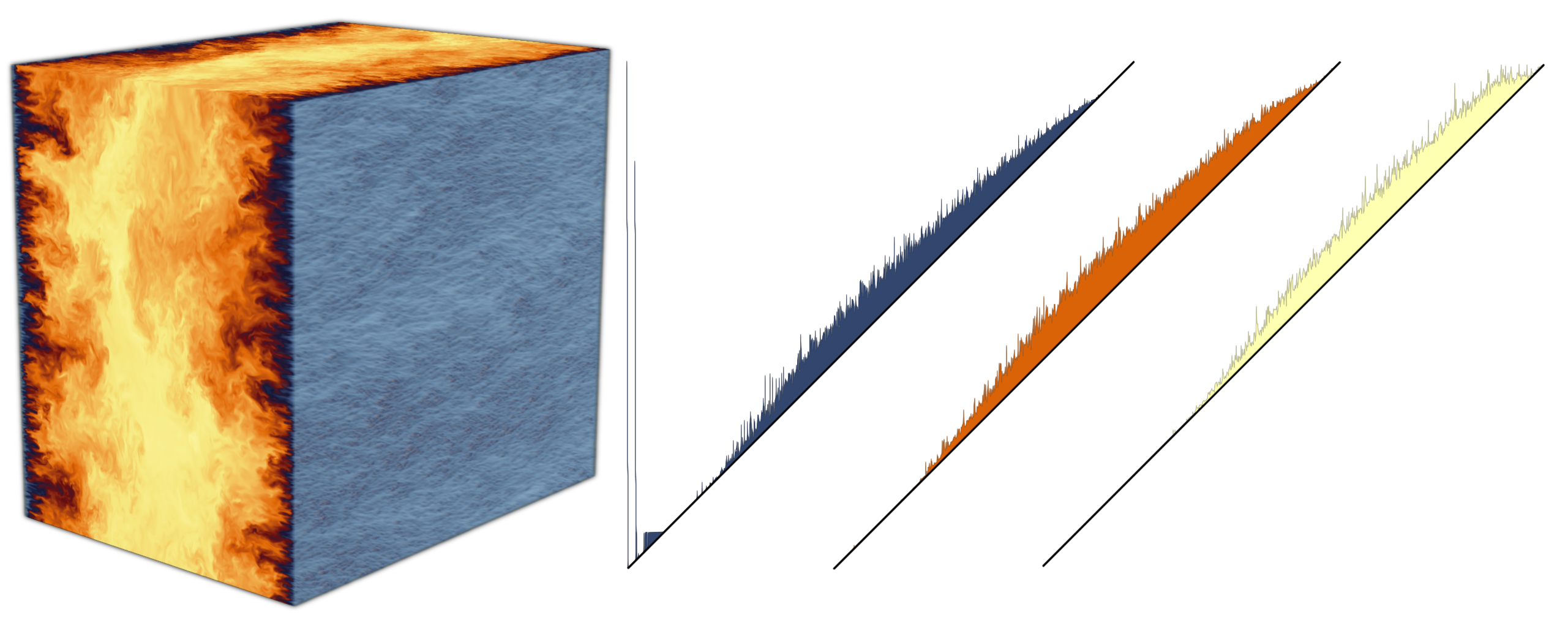}
    \caption{Persistence diagram on a subset of Turbulent Channel Flow (2048x1920x1536, 
    6 billion vertices).
    The dataset (left) is the pressure field of a direct simulation of a turbulent flow in a 
    plane channel. The execution was performed on 16 nodes of 32 cores and
256GB
    \julien{of RAM}
    each \julien{(512 cores and 4096GB of RAM total).}
    The persistence diagram (right: $\diagram_{0}$ in blue, $\diagram_{1}$ in
orange,
     $\diagram_{2}$ in yellow) was computed in 174 seconds
     (19 million pairs).}
%      and
%     contains 19 million pairs ().}
    \label{fig_dns}
\end{figure}

We ran DDMS on a larger dataset (Turbulent Channel Flow \cite{openSciVisDataSets}) to show 
our algorithm's capability to handle massive datasets. This dataset is a direct 
simulation of a fully developed flow at different Reynolds numbers in a
plane channel. The scalar field is the three dimensional pressure field and has been 
converted to single-precision floating-point numbers for a lower 
memory consumption.
%% (thereby reducing memory consumption at runtime).
The computation was run on a subset of the original dataset of size $2048$x$1920$x$1536$
which is approximately 6 billion vertices. The memory bottleneck of our implementation
mainly lies in the computation of the discrete gradient, which is the
most memory-consuming step. For this example, we used 
TTK's compile option $\verb|TTK_ENABLE_DCG_OPTIMIZE_MEMORY|$
\julien{which optimizes the gradient memory footprint (by trading on its
computation time).}
% that optimizes memory
% in the gradient to reduce its memory footprint but slows down its computation.
The execution was performed on 16 nodes of 32 cores and 256GB \julien{of
RAM} each
\julien{(512 cores and 4096GB of RAM total).}
% for a total of 512 cores and 4096GB of memory.
The persistence diagram shown in \autoref{fig_dns} was computed in 174 seconds 
and contained a little under 19 million pairs.
\julien{For comparison, out-of-core techniques \cite{Wagner23} require
several hours of computation for datasets of similar size, which further
illustrates the practical interest of our work for high-performance
computing contexts.}

\subsection{Limitations}

For completeness, we discuss here some limitations of our work.
Our
\julien{implementation of the}
DDMS algorithm only supports structured grids at the time of writing this
paper,
even though TTK support both structured and unstructured grids.
\julien{Although the original DMS implementation has been validated on both,
additional tests would be required to validate our DDMS implementation on
unstructured grids.} \majorRevision{Certain aspects of practical use may be 
different when using unstructured meshes. For example, as unstructured grids 
require more memory, our tests and benchmarks may need some modifications. 
The \pierreRevision{performance} %execution
may \pierreRevision{also} be more memory-bound. Substantial additional
tests are \pierreRevision{thus} required to evaluate
the performance of our algorithm on such grids.}
% dditional tests would be required to validate }
% Unstructured grids would require additional processing, in particular for detecting
% the infinite pairs of datasets with more than one connected component. This has
% not been implemented yet.
Additionally, several specialized domain representations
popular in scientific computing such as adaptive mesh refinement (AMR) are not supported 
by TTK and therefore by our
\julien{work.}
% algorithm.

The primary memory  bottleneck of our implementation lies in the computation of the
gradient. Indeed, this computation is not only costly in terms of execution time but also
in terms of memory footprint, as discrete vectors are computed for all simplices of all 
dimensions. This generates a significant memory footprint, \julien{which may be
addressed in the future by an improved compact encoding of the
discrete vectors, exploiting the regular structures of structured grids}.

\julien{Finally, similarly to DMS, since our work also exploits Robins'
discrete gradient \cite{robins_pami11}, it suffers from the same limitation
regarding an extension to higher dimensions. In particular, strong guarantees
on the restriction of critical cells upon homology changes of the
sub-complexes are provided in up to three dimensions. Beyond, additional,
spurious critical cells may appear, limiting the effectiveness of our approach.}

\section{Conclusion}

This paper introduced a new algorithm for the \julien{efficient} computation of
the persistence diagram
on scalar data in a distributed-memory setting using a hybrid MPI+thread 
parallelization.
% Our work builds upon an existing algorithm, the Discrete Morse
% Sandwich (DMS) \cite{guillou_tech22}, already partly parallelized for shared-memory.
% Similarly to DMS, DDMS relies on a stratification strategy. First, the discrete
% gradient is computed. This step is costly but embarrassingly
% parallel. %% Pierre : "parallelize" se s'emploie pas trop je crois %trivially parallelizable.
% \pierre{P2E: il vaut mieux à partir d'ici rappeler (brièvement) tes contributions
%   que réexpliquer DMS/DDMS}
% Then $\diagram_{0}$ and $\diagram_{d-1}$ are computed by processing stable and
% unstable sets, extracting a simplified graph from the sets and applying to it a
% self-correcting algorithm to produce the pairs. $\diagram_{1}$ is then computed
% from the unpaired 1- and 2-saddles by homologous propagation.
The performance of our algorithm was tested on a set of 8 datasets representing 
various use cases.
Thanks to our algorithmic improvements for $\diagram_{0}$, 
$\diagram_{1}$ and $\diagram_{2}$, and to the scalability of the
discrete gradient computation, we have shown that our
\julien{approach}
% implementation
delivers significant speedups \julien{on}  up to 512 cores, with parallel
efficiencies up to 50\% in strong scaling and up to 80\% in weak scaling.
%%
%% For both strong and weak scaling, our implementation showed good 
%% efficiency for most datasets, in the range of 20\% to 50\% for strong scaling and 
%% 35\% to 80\% for weak scaling, successfully harnessing the scalability of the discrete
%% gradient.
DDMS also outperforms \majorRevision{\emph{DIPHA}},
% \julien{to our knowledge}
the only publicly available implementation
for computing the persistence diagram in %a
distributed memory, % setting,
by a factor of
\julien{$\times 8$}
% 8x
on average.
%% DDMS is on average 8 times faster than \majorRevision{\emph{DIPHA}}.
It produces a slightly larger %bigger
memory footprint than \majorRevision{\emph{DIPHA}} on few nodes, but a
\julien{smaller}
% lower
one
%outperforms \majorRevision{\emph{DIPHA}} : pierre : outperform fait plutot penser à
%de la performance qu'à de la consommation mémoire 
on 16 %a larger number of
nodes.

Additionally, we showed that our algorithm is capable of processing massive datasets
by running DDMS on a larger dataset (Turbulent Channel Flow \cite{openSciVisDataSets})
of 6 billion vertices in under
% 4
\julien{3}
minutes on 512 cores.

In the future, we will investigate strategies to reduce the memory footprint 
of the discrete gradient. Indeed, as previously stated, the primary memory 
bottleneck of our implementation lies in the computation of the gradient. 
Reducing its memory footprint while maintaining its execution
speed would be beneficial for DDMS. 
An avenue for further research is \pierre{deploying} %parallelizing
\eve{DDMS 
on heterogenous architectures.
Given the gradient algorithm's embarrassingly parallel %%and
%%compute-intensive %% Pierre : compute-intensive indique un bonne
%%intensité arithmétique ce qui ne doit pas être le cas il me semble pour le calcul
%%du gradient discret 
nature, 
\pierre{CPU architectures} with integrated GPUs, such as \pierre{the} AMD %Pierre 's
Instinct MI300A APU, 
may further accelerate DDMS by leveraging GPU 
cores for \pierre{the} gradient computation and CPU cores for the computation of $\diagram_{0}$, 
$\diagram_{1}$ and $\diagram_{2}$.}

%% has not 
%% been explored here as it is outside of the scope of this paper. However, such 
%% optimization could offer a potential approach for improving the memory efficiency of DDMS.

Another research direction is out-of-core computation, which enables processing
large datasets on a single compute server by loading data in and out of memory.
\julien{Specifically, extensions of our
self-correcting pairing (for $\diagram_{0}$ and $\diagram_{2}$,
\autoref{sec_selfcorrecting}) as well as our
anticipation strategy for homologous
propagation (for $\diagram_{1}$, \autoref{sec_anticipation}) could be considered
for out-of-core contexts, but significant adaptations would be required to
maintain the time performance of our approach.}
% While memory is not distributed, this approach resembles distributed-memory systems,
% therefore DDMS could be build upon to produce an out-of-core approach for
% persistence diagram computation.
%Apply stratification strategy to other representations like Contour trees?

\section*{Acknowledgments}
% \Acknowledgments{
\footnotesize{
This work is partially supported by the European Commission grant
ERC-2019-COG \emph{``TORI''} (ref. 863464, \url{https://erc-tori.github.io/}).}
%
%
% \pierre{[uniformiser le nom des journaux/confs : sigle pour la
%     plupart, mais pas pour tous]}
\bibliographystyle{IEEEtran}
\bibliography{paper}

\begin{thebibliography}{10}
\providecommand{\url}[1]{#1}
\csname url@samestyle\endcsname
\providecommand{\newblock}{\relax}
\providecommand{\bibinfo}[2]{#2}
\providecommand{\BIBentrySTDinterwordspacing}{\spaceskip=0pt\relax}
\providecommand{\BIBentryALTinterwordstretchfactor}{4}
\providecommand{\BIBentryALTinterwordspacing}{\spaceskip=\fontdimen2\font plus
\BIBentryALTinterwordstretchfactor\fontdimen3\font minus
  \fontdimen4\font\relax}
\providecommand{\BIBforeignlanguage}[2]{{%
\expandafter\ifx\csname l@#1\endcsname\relax
\typeout{** WARNING: IEEEtran.bst: No hyphenation pattern has been}%
\typeout{** loaded for the language `#1'. Using the pattern for}%
\typeout{** the default language instead.}%
\else
\language=\csname l@#1\endcsname
\fi
#2}}
\providecommand{\BIBdecl}{\relax}
\BIBdecl

\bibitem{laney_vis06}
D.~E. Laney, P.~Bremer, A.~Mascarenhas, P.~Miller, and V.~Pascucci,
  ``{Understanding the Structure of the Turbulent Mixing Layer in Hydrodynamic
  Instabilities},'' \emph{IEEE Trans. Vis. and Comp. Graph. (Proc. of IEEE
  VIS)}, 2006.

\bibitem{bremer_tvcg11}
P.~Bremer, G.~Weber, J.~Tierny, V.~Pascucci, M.~Day, and J.~Bell,
  ``{Interactive Exploration and Analysis of Large Scale Simulations Using
  Topology-based Data Segmentation},'' \emph{IEEE Trans. Vis. and Comp.
  Graph.}, 2011.

\bibitem{gyulassy_ev14}
A.~Gyulassy, P.~Bremer, R.~Grout, H.~Kolla, J.~Chen, and V.~Pascucci,
  ``{Stability of Dissipation Elements: A case study in combustion},''
  \emph{EuroVis: Proc. of Eurographics Conference on Visualization}, 2014.

\bibitem{kasten_tvcg11}
J.~Kasten, J.~Reininghaus, I.~Hotz, and H.~Hege, ``{Two-Dimensional
  Time-Dependent Vortex Regions Based on the Acceleration Magnitude},''
  \emph{IEEE Trans. Vis. and Comp. Graph.}, 2011.

\bibitem{nauleau_ldav22}
F.~Nauleau, F.~Vivodtzev, T.~Bridel-Bertomeu, H.~Beaugendre, and J.~Tierny,
  ``{Topological Analysis of Ensembles of Hydrodynamic Turbulent Flows -- An
  Experimental Study},'' in \emph{IEEE Symposium on Large Data Analysis and
  Visualization}, 2022.

\bibitem{beiNuclear16}
D.~Maljovec, B.~Wang, P.~Rosen, A.~Alfonsi, G.~Pastore, C.~Rabiti, and
  V.~Pascucci, ``{Topology-Inspired Partition-Based Sensitivity Analysis and
  Visualization of Nuclear Simulations},'' in \emph{IEEE Pacific Vis}, 2016.

\bibitem{ChazalGOS13}
F.~Chazal, L.~J. Guibas, S.~Y. Oudot, and P.~Skraba, ``{Persistence-Based
  Clustering in Riemannian Manifolds},'' \emph{Journal of the {ACM}}, 2013.

\bibitem{topoMap}
H.~Doraiswamy, J.~Tierny, P.~J.~S. Silva, L.~G. Nonato, and C.~T. Silva,
  ``{TopoMap: {A} 0-dimensional Homology Preserving Projection of
  High-Dimensional Data},'' \emph{IEEE Trans. Vis. and Comp. Graph. (Proc. of
  IEEE VIS)}, 2020.

\bibitem{harshChemistry}
H.~Bhatia, A.~G. Gyulassy, V.~Lordi, J.~E. Pask, V.~Pascucci, and P.-T.
  Bremer., ``{TopoMS: Comprehensive Topological Exploration for Molecular and
  Condensed-Matter Systems},'' \emph{Journal of Computational Chemistry}, 2018.

\bibitem{chemistry_vis14}
D.~Guenther, R.~Alvarez-Boto, J.~Contreras-Garcia, J.-P. Piquemal, and
  J.~Tierny, ``{Characterizing Molecular Interactions in Chemical Systems},''
  \emph{IEEE Trans. Vis. and Comp. Graph. (Proc. of IEEE VIS)}, 2014.

\bibitem{Malgorzata19}
M.~Olejniczak, A.~S.~P. Gomes, and J.~Tierny, ``{A Topological Data Analysis
  Perspective on Non-Covalent Interactions in Relativistic Calculations},''
  \emph{International Journal of Quantum Chemistry}, 2019.

\bibitem{sousbie11}
T.~Sousbie, ``{The Persistent Cosmic Web and its Filamentary Structure: Theory
  and Implementations},'' \emph{{Royal Astronomical Society}}, 2011.

\bibitem{shivashankar2016felix}
N.~Shivashankar, P.~Pranav, V.~Natarajan, R.~van~de Weygaert, E.~P. Bos, and
  S.~Rieder, ``Felix: A topology based framework for visual exploration of
  cosmic filaments,'' \emph{IEEE Trans. Vis. and Comp. Graph.}, 2016.

\bibitem{edelsbrunner09}
H.~Edelsbrunner and J.~Harer, \emph{Computational Topology: An
  Introduction}.\hskip 1em plus 0.5em minus 0.4em\relax American Mathematical
  Society, 2009.

\bibitem{heine16}
C.~Heine, H.~Leitte, M.~Hlawitschka, F.~Iuricich, L.~De~Floriani,
  G.~Scheuermann, H.~Hagen, and C.~Garth, ``{A Survey of Topology-based Methods
  in Visualization},'' \emph{Computer Graphics Journal}, 2016.

\bibitem{edelsbrunner02}
H.~Edelsbrunner, D.~Letscher, and A.~Zomorodian, ``Topological persistence and
  simplification,'' \emph{Discrete and Computational Geometry}, 2002.

\bibitem{Parsa12}
S.~Parsa, ``{A Deterministic \emph{o(m log m)} Time Algorithm for the Reeb
  Graph},'' in \emph{Symposium on Computational Geometry}, 2012.

\bibitem{gueunet_egpgv19}
C.~Gueunet, P.~Fortin, J.~Jomier, and J.~Tierny, ``{Task-based Augmented Reeb
  Graphs with Dynamic ST-Trees},'' in \emph{{Eurographics Symposium on Parallel
  Graphics and Visualization}}, 2019.

\bibitem{tarasov98}
S.~Tarasov and M.~Vyali, ``{Construction of contour trees in 3D in O(n log n)
  steps},'' in \emph{Symposium on Computational Geometry}, 1998.

\bibitem{carr00}
H.~Carr, J.~Snoeyink, and U.~Axen, ``Computing contour trees in all
  dimensions,'' in \emph{Symp. on Dis. Alg.}, 2000.

\bibitem{gueunet_tpds19}
C.~Gueunet, P.~Fortin, J.~Jomier, and J.~Tierny, ``{Task-Based Augmented
  Contour Trees with Fibonacci Heaps},'' \emph{{IEEE} {Trans. Parallel Distrib.
  Syst.}}, 2019.

\bibitem{BremerEHP04}
P.~Bremer, H.~Edelsbrunner, B.~Hamann, and V.~Pascucci, ``{A Topological
  Hierarchy for Functions on Triangulated Surfaces},'' \emph{IEEE Trans. Vis.
  and Comp. Graph.}, 2004.

\bibitem{GyulassyNPBH06}
A.~Gyulassy, V.~Natarajan, V.~Pascucci, P.~Bremer, and B.~Hamann, ``{A
  Topological Approach to Simplification of Three-Dimensional Scalar
  Functions},'' \emph{IEEE Trans. Vis. and Comp. Graph.}, 2006.

\bibitem{Defl15}
L.~De~Floriani, U.~Fugacci, F.~Iuricich, and P.~Magillo, ``Morse complexes for
  shape segmentation and homological analysis: discrete models and
  algorithms,'' \emph{Computer Graphics Journal}, 2015.

\bibitem{gyulassy_vis18}
A.~Gyulassy, P.~Bremer, and V.~Pascucci, ``{Shared-Memory Parallel Computation
  of Morse-Smale Complexes with Improved Accuracy},'' \emph{IEEE Trans. Vis.
  and Comp. Graph. (Proc. of IEEE VIS)}, 2018.

\bibitem{tierny_vis12}
J.~Tierny and V.~Pascucci, ``Generalized topological simplification of scalar
  fields on surfaces,'' \emph{IEEE Trans. Vis. and Comp. Graph. (Proc. of IEEE
  VIS)}, 2012.

\bibitem{Lukasczyk_vis20}
J.~Lukasczyk, C.~Garth, R.~Maciejewski, and J.~Tierny, ``{Localized Topological
  Simplification of Scalar Data},'' \emph{IEEE Trans. Vis. and Comp. Graph.
  (Proc. of IEEE VIS)}, 2020.

\bibitem{carr04}
H.~A. Carr, J.~Snoeyink, and M.~van~de Panne, ``{Simplifying Flexible
  Isosurfaces Using Local Geometric Measures},'' in \emph{IEEE VIS}, 2004.

\bibitem{topoAngler}
A.~Bock, H.~Doraiswamy, A.~Summers, and C.~T. Silva, ``{TopoAngler: Interactive
  Topology-Based Extraction of Fishes},'' \emph{IEEE Trans. Vis. and Comp.
  Graph. (Proc. of IEEE VIS)}, 2018.

\bibitem{soler_pv18}
M.~Soler, M.~Plainchault, B.~Conche, and J.~Tierny, ``{Topologically Controlled
  Lossy Compression},'' in \emph{IEEE Pacific Vis}, 2018.

\bibitem{soler_ldav18}
------, ``{Lifted {W}asserstein Matcher for Fast and Robust Topology
  Tracking},'' in \emph{IEEE Symposium on Large Data Analysis and
  Visualization}, 2018.

\bibitem{soler_ldav19}
M.~Soler, M.~Petitfrere, G.~Darche, M.~Plainchault, B.~Conche, and J.~Tierny,
  ``{Ranking Viscous Finger Simulations to an Acquired Ground Truth with
  Topology-Aware Matchings},'' in \emph{IEEE Symposium on Large Data Analysis
  and Visualization}, 2019.

\bibitem{LukasczykGWBML20}
J.~Lukasczyk, C.~Garth, G.~H. Weber, T.~Biedert, R.~Maciejewski, and H.~Leitte,
  ``{Dynamic Nested Tracking Graphs},'' \emph{IEEE Trans. Vis. and Comp. Graph.
  (Proc. of IEEE VIS)}, 2019.

\bibitem{favelier_vis18}
G.~Favelier, N.~Faraj, B.~Summa, and J.~Tierny, ``{Persistence Atlas for
  Critical Point Variability in Ensembles},'' \emph{IEEE Trans. Vis. and Comp.
  Graph. (Proc. of IEEE VIS)}, 2018.

\bibitem{vidal_vis19}
J.~Vidal, J.~Budin, and J.~Tierny, ``{Progressive Wasserstein Barycenters of
  Persistence Diagrams},'' \emph{IEEE Trans. Vis. and Comp. Graph. (Proc. of
  IEEE VIS)}, 2019.

\bibitem{KontakVT19}
M.~Kontak, J.~Vidal, and J.~Tierny, ``{Statistical Parameter Selection for
  Clustering Persistence Diagrams},'' in \emph{2019 {IEEE/ACM} UrgentHPC@SC},
  2019.

\bibitem{guillou_tech22}
P.~Guillou, J.~Vidal, and J.~Tierny, ``{Discrete Morse Sandwich: Fast
  Computation of Persistence Diagrams for Scalar Data -- An Algorithm and A
  Benchmark},'' \emph{IEEE Trans. Vis. and Comp. Graph.}, 2023.

\bibitem{ttk19}
T.~Bin~Masood, J.~Budin, M.~Falk, G.~Favelier, C.~Garth, C.~Gueunet,
  P.~Guillou, L.~Hofmann, P.~Hristov, A.~Kamakshidasan, C.~Kappe, P.~Klacansky,
  P.~Laurin, J.~Levine, J.~Lukasczyk, D.~Sakurai, M.~Soler, P.~Steneteg,
  J.~Tierny, W.~Usher, J.~Vidal, and M.~Wozniak, ``{An Overview of the Topology
  ToolKit},'' in \emph{TopoInVis}, 2019.

\bibitem{ttk17}
J.~Tierny, G.~Favelier, J.~A. Levine, C.~Gueunet, and M.~Michaux, ``The
  {T}opology {T}ool{K}it,'' \emph{IEEE Trans. Vis. and Comp. Graph. (Proc. of
  IEEE VIS)}, 2017, \url{https://topology-tool-kit.github.io/}.

\bibitem{leguillou_tech24}
E.~Le$\hspace{3pt}$Guillou, M.~Will, P.~Guillou, J.~Lukasczyk, P.~Fortin,
  C.~Garth, and J.~Tierny, ``{TTK is Getting MPI-Ready},'' \emph{IEEE Trans.
  Vis. and Comp. Graph.}, 2024.

\bibitem{dipha}
U.~Bauer, M.~Kerber, and J.~Reininghaus, ``{Distributed Computation of
  Persistent Homology},'' in \emph{Proceedings of the Workshop on Algorithm
  Engineering and Experiments}, 2014.

\bibitem{edelsbrunner90}
H.~Edelsbrunner and E.~P. Mucke, ``{Simulation of Simplicity: A Technique to
  Cope with Degenerate Cases in Geometric Algorithms},'' \emph{ACM Trans. on
  Graphics}, 1990.

\bibitem{freudenthal42}
{H. Freudenthal}, ``{Simplizialzerlegungen von beschrankter Flachheit},''
  \emph{{Annals of Mathematics}}, vol.~{43}, pp. 580--582, 1942.

\bibitem{paraviewBook}
J.~Ahrens, B.~Geveci, and C.~Law, ``{ParaView: An End-User Tool for Large-Data
  Visualization},'' \emph{The Visualization Handbook}, 2005.

\bibitem{forman98}
R.~Forman, ``{A User's Guide to Discrete Morse Theory},'' \emph{AM}, 1998.

\bibitem{milnor63}
J.~Milnor, \emph{{Morse Theory}}.\hskip 1em plus 0.5em minus 0.4em\relax
  {Princeton University Press}, 1963.

\bibitem{morseQuote}
M.~Morse, ``{The Calculus of Variations in the Large},'' in \emph{American
  Mathematical Society}, 1934.

\bibitem{robins_pami11}
V.~Robins, P.~J. Wood, and A.~P. Sheppard, ``{Theory and Algorithms for
  Constructing Discrete Morse Complexes from Grayscale Digital Images},''
  \emph{{IEEE Trans. Pattern Anal. Mach. Intell.}}, 2011.

\bibitem{Nigmetov20}
D.~Morozov and A.~Nigmetov, ``{Brief Announcement: Towards Lockfree Persistent
  Homology},'' in \emph{{Symposium on Parallelism in Algorithms and
  Architectures}}, 2020.

\bibitem{barannikovFramedMorseComplex1994}
{S. A. Barannikov}, ``{The Framed {{Morse}} Complex and Its Invariants},'' in
  \emph{{{ADVSOV}}}.\hskip 1em plus 0.5em minus 0.4em\relax American
  Mathematical Society, 1994.

\bibitem{frosini99}
P.~Frosini and C.~Landi, ``{Size Theory as a Topological Tool for Computer
  Vision},'' \emph{Pattern Recognition and Image Analysis}, 1999.

\bibitem{robins99}
V.~Robins, ``{Toward Computing Homology from Finite Approximations},''
  \emph{Topology Proceedings}, 1999.

\bibitem{perseusPaper}
K.~Mischaikow and V.~Nanda, ``{Morse Theory for Filtrations and Efficient
  Computation of Persistent Homology},'' \emph{Discrete and Computational
  Geometry}, 2012.

\bibitem{ripser}
U.~Bauer, ``{Ripser: efficient computation of Vietoris-Rips persistence
  barcodes},'' \url{https://github.com/Ripser/ripser/}, 2019.

\bibitem{bauer2013clearcompresscomputingpersistent}
U.~Bauer, M.~Kerber, and J.~Reininghaus, ``{Clear and Compress: Computing
  Persistent Homology in Chunks},'' in \emph{Topological Methods in Data
  Analysis and Visualization III}.\hskip 1em plus 0.5em minus 0.4em\relax
  Springer International Publishing, 2014, pp. 103--117.

\bibitem{GuntherRWH12}
D.~G{\"{u}}nther, J.~Reininghaus, H.~Wagner, and I.~Hotz, ``{Efficient
  computation of 3D Morse-Smale complexes and persistent homology using
  discrete Morse theory},'' \emph{{The Visual Computer}}, 2012.

\bibitem{iuricich21}
F.~Iuricich, ``Persistence cycles for visual exploration of persistent
  homology,'' \emph{IEEE Trans. Vis. and Comp. Graph.}, 2021.

\bibitem{Wagner23}
H.~Wagner, ``{Slice, Simplify and Stitch: Topology-Preserving Simplification
  Scheme for Massive Voxel Data},'' in \emph{Symposium on Computational
  Geometry}, 2023.

\bibitem{BauerKRW17}
U.~Bauer, M.~Kerber, J.~Reininghaus, and H.~Wagner, ``{Phat - Persistent
  Homology Algorithms Toolbox},'' \emph{Journal of Symbolic Computation}, 2017.

\bibitem{gudhi}
C.~Maria, J.~Boissonnat, M.~Glisse, and M.~Yvinec, ``{The Gudhi Library:
  Simplicial Complexes and Persistent Homology},'' in \emph{{Mathematical
  Software}}, 2014, \url{https://github.com/GUDHI/}.

\bibitem{eirene}
G.~Henselman-Petrusek, ``{Eirene.jl package for Homological Algebra},''
  \url{https://github.com/Eetion/Eirene.jl}, 2018.

\bibitem{toth2025user}
C.~T{\'o}th, D.~J.~D. Cruz, and H.~Oberhauser, ``A user's guide to
  \texttt{KSig}: Gpu-accelerated computation of the signature kernel,''
  \emph{arXiv preprint arXiv:2501.07145}, 2025.

\bibitem{oberhauser2023}
D.~Lee and H.~Oberhauser, ``{The Signature Kernel},'' \emph{arXiv preprint
  arXiv:2305.04625}, 2023.

\bibitem{otterRoadmapComputationPersistent2017}
N.~Otter, M.~Porter, U.~Tillmann, P.~Grindrod, and H.~Harrington, ``A roadmap
  for the computation of persistent homology,'' \emph{EPJ Data Science},
  vol.~6, p.~17, 2017.

\bibitem{ceccaroniDistributedApproachPersistent2024a}
R.~Ceccaroni, L.~Di~Rocco, U.~Ferraro~Petrillo, and P.~Brutti, ``{A Distributed
  Approach for Persistent Homology Computation on a Large Scale},''
  \emph{Journal of Symbolic Computation}, 2024.

\bibitem{cadmus}
\BIBentryALTinterwordspacing
A.~Nigmetov and D.~Morozov, ``{Distributed Computation of Persistent
  Cohomology},'' 2024. [Online]. Available:
  \url{https://arxiv.org/abs/2410.16553}
\BIBentrySTDinterwordspacing

\bibitem{psort}
\BIBentryALTinterwordspacing
D.~Cheng, V.~Shah, J.~Gilbert, and A.~Edelman, ``A novel parallel sorting
  algorithm for contemporary architectures,'' 01 2007. [Online]. Available:
  \url{https://github.com/DIPHA/dipha/tree/master/externals/psort-1.0}
\BIBentrySTDinterwordspacing

\bibitem{smirnov17}
D.~Smirnov and D.~Morozov, ``{Triplet {M}erge {T}rees},'' in \emph{TopoInVis},
  2017.

\bibitem{denisMPIOverlapBenchmark2016a}
A.~Denis and F.~Trahay, ``{{MPI Overlap}}: {{Benchmark}} and {{Analysis}},'' in
  \emph{{{International Conference}} on {{Parallel Processing}}}, 2016.

\bibitem{hager2011}
G.~Hager, G.~Schubert, T.~Schoenemeyer, and G.~Wellein, ``Prospects for
  {{Truly}} {{Asynchronous}} {{Communication}} with {{Pure}} {{MPI}} and
  {{Hybrid}} {{MPI/OpenMP}} on {{Current}} {{Supercomputing}} {{Platforms}},''
  in \emph{Proceedings of The Cray User Group}, 2011.

\bibitem{openSciVisDataSets}
P.~Klacansky, ``{Open Scientific Visualization Data Sets},''
  \\\url{https://klacansky.com/open-scivis-datasets/}, 2020.

\bibitem{ttkData}
{TTK Contributors}, ``{TTK Data},''
  \\\url{https://github.com/topology-tool-kit/ttk-data/tree/dev}, 2020.

\end{thebibliography}

\section{Biography Section}
 
\vspace{-2ex}
\begin{IEEEbiography}[{\includegraphics[width=1in,height=1.25in,clip,
  keepaspectratio]{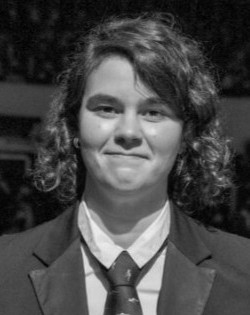}}]{Eve Le Guillou}
  is a Ph.D. student at Sorbonne Université. She received a M.S. degree in 2020 in Computer Science
  from Cranfield University as well as an engineering degree in 2021 from École Centrale de Lille.
  She is an active contributor to the Topology ToolKit (TTK), an open source library for
  topological data analysis. Her notable contributions to TTK include the port to MPI
  of TTK's data structure and of several algorithms.
  \end{IEEEbiography}

  \vspace{-2ex}
  \begin{IEEEbiography}[{\includegraphics[width=1in,height=1.25in,
    clip,keepaspectratio]{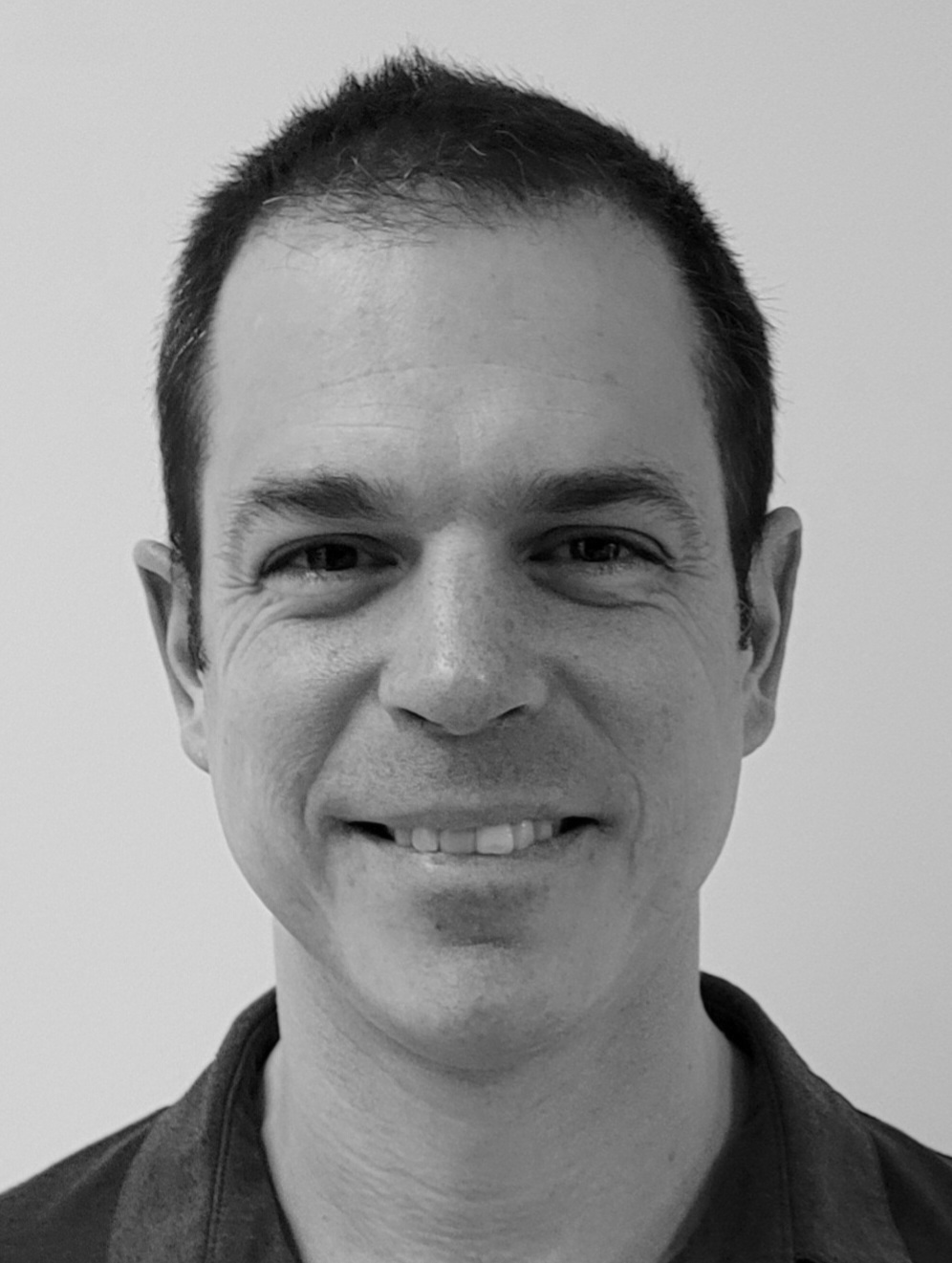}}]{Pierre Fortin}
received his Ph.D. degree in Computer Science from the University
of Bordeaux in 2006.
He was an Assistant Professor at Sorbonne
University from 2007, and he is Professor
at University of Lille since 2023. His research interests include parallel
algorithmic and high performance scientific computing.
\end{IEEEbiography}

\vspace{-2ex}
\begin{IEEEbiography}[{\includegraphics[width=1in,height=1.25in,clip,
  keepaspectratio]{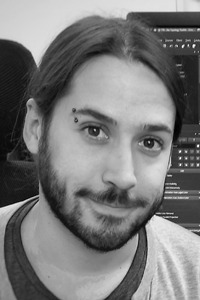}}]{Julien Tierny}
  received the Ph.D. degree in Computer Science from the University
  of Lille in 2008. He is a CNRS research director at
  Sorbonne University. Prior to his CNRS tenure, he held a
  Fulbright fellowship and was a post-doctoral
  researcher at the University of Utah.
  His research expertise lies in topological methods for data analysis
  and visualization.
  He is the founder and lead developer of the Topology ToolKit
  (TTK), an open source library for topological data analysis.
  \end{IEEEbiography}

\vfill

\end{document}